\documentclass[aps,pra,reprint,superscriptaddress,floatfix,longbibliography]{revtex4-1}

\usepackage{amsmath}    
\usepackage{amssymb}
\usepackage{graphicx}   
\usepackage[lofdepth,lotdepth, caption=false]{subfig}
\usepackage{verbatim}   
\usepackage{color}      
\usepackage{hyperref}   
\usepackage{dcolumn}

\def\lnod{La$_2$NiO$_{4+\delta}$}
\def\lsno{La$_{2-x}$Sr$_x$NiO$_4$}

\def\bscco{Bi$_2$Sr$_2$CaCu$_2$O$_{8+\delta}$}
\def\newr{\color{black}}

\begin{document}

\newcommand{\LSNO}{La$_{2-x}$Sr$_{x}$NiO$_{4}$}
\newcommand{\LSNOn}{La$_{1.75}$Sr$_{0.25}$NiO$_{4}$}

\title{Evidence for a nematic phase in \LSNOn}

\author{Ruidan~Zhong}
\affiliation{Condensed Matter Physics and Materials Science Department, Brookhaven National Laboratory, Upton, NY 11973, USA}
\affiliation{Materials Science and Engineering Department, Stony Brook University, Stony Brook, NY 11794, USA}
\author{Barry~L.~Winn}
\affiliation{Quantum Condensed Matter Division, Oak Ridge National Laboratory, Oak Ridge, Tennessee 37831, USA}
\author{Genda Gu}
\affiliation{Condensed Matter Physics and Materials Science Department, Brookhaven National Laboratory, Upton, NY 11973, USA}
\author{Dmitry Reznik}
\affiliation{Department of Physics, University of Colorado, Boulder, CO 80304, USA}
\author{J.~M.~Tranquada}
\email{jtran@bnl.gov}
\affiliation{Condensed Matter Physics and Materials Science Department, Brookhaven National Laboratory, Upton, NY 11973, USA}

\date{April 28, 2017}

\begin{abstract}
Determining the nature of electronic states in doped Mott insulators remains a challenging task.  In the case of tetragonal  La$_{2-x}$Sr$_{x}$NiO$_{4}$, the occurrence of diagonal charge and spin stripe order in the ground state is now well established.  In contrast, the nature of the high-temperature ``disordered'' state from which the stripe order develops has long been a subject of controversy, with considerable speculation regarding a polaronic liquid.  Following on the recent detection of dynamic charge stripes, we use neutron scattering measurements on an $x=0.25$ crystal to demonstrate that the dispersion of the charge stripe excitations is anisotropic.   This observation provides compelling evidence for the presence of electronic nematic order.
\end{abstract}

\pacs{PACS: 71.45.Lr, 71.27.+a, 78.70.Nx}

\maketitle

Doping holes into the the correlated insulator La$_2$NiO$_4$ leads to the destruction of commensurate antiferromagnetic order and the formation of charge and spin stripes \cite{tran94a,lee97,yosh00,ulbr12b}.  With the recent evidence for ubiquitous charge-density-wave (CDW) order in cuprate superconductors \cite{keim15,comi16}, the problem of understanding CDWs in correlated systems continues to receive considerable attention.  While there are important differences between the $S=\frac12$ cuprates and the $S=1$ nickelates, explaining even the nonsuperconducting systems remains a challenge.  There has been considerable experimental work characterizing the spin and charge order of \lnod\ and \lsno \cite{brom03,ulbr12b,tran13a}, and  theoretical calculations have captured various aspects of the ground state \cite{zaan94,hott04,racz06,yama07,schw08}.  The situation is much less clear when one considers the state from which the stripes develop on cooling.

On melting the charge stripes in a nickelate sample, the conductivity increases \cite{kats96,home03,cosl13}; however, these materials are, at best, ``bad metals'' \cite{emer95b}, lacking coherent quasiparticle states.  Optical conductivity measurements confirm the absence of a Drude peak, a standard signature of coherent conduction \cite{ido91,kats96,home03,cosl13}.  Disorder does not play a significant role, as the situation is nearly identical even when the dopant ions have long-range order, as in La$_2$NiO$_{4.133}$ \cite{home03}.  Given these conditions, the standard picture of charge-density-wave order is completely inapplicable, as it is based on a model of nearly-free electrons interacting with the lattice \cite{berl79,grun88}.   Even if it were, there is no model for a state of fluctuating CDWs in two or more dimensions \cite{lee73}.  Discussions of the disordered state have generally invoked a liquid of polarons \cite{anis92,chen93,fehs95,baum98,masc16}

In this Letter, we present inelastic neutron scattering measurements characterizing charge-stripe fluctuations at temperatures above the spin-ordering transition in \lsno\ (LSNO) with $x=0.25$.  The existence of such fluctuations in LSNO with $x=0.33$ has been inferred from a study of the temperature dependence of Debye-Waller factors \cite{abey13}, and the first direct evidence was obtained by Anissimova {\it et al.}\ \cite{anis14}.   In the present case, we study a regime where the correlation lengths associated with charge stripes are short, so that there is no long-range translational symmetry breaking.  At the same time, we show that the dispersion of the charge-stripe fluctuations is anisotropic, with a lower effective velocity along the modulation direction, comparable to that of transverse acoustic phonons.  This anisotropy establishes that the electronic rotational symmetry within the NiO$_2$ planes is reduced to $C_2$, whereas the atomic structure retains $C_4$ symmetry.  Hence, the high-temperature electronic phase appears to have nematic order \cite{kive98,frad10,nie14}.

\begin{figure*}[ht]
 \centering
    \includegraphics[width=1.5\columnwidth]{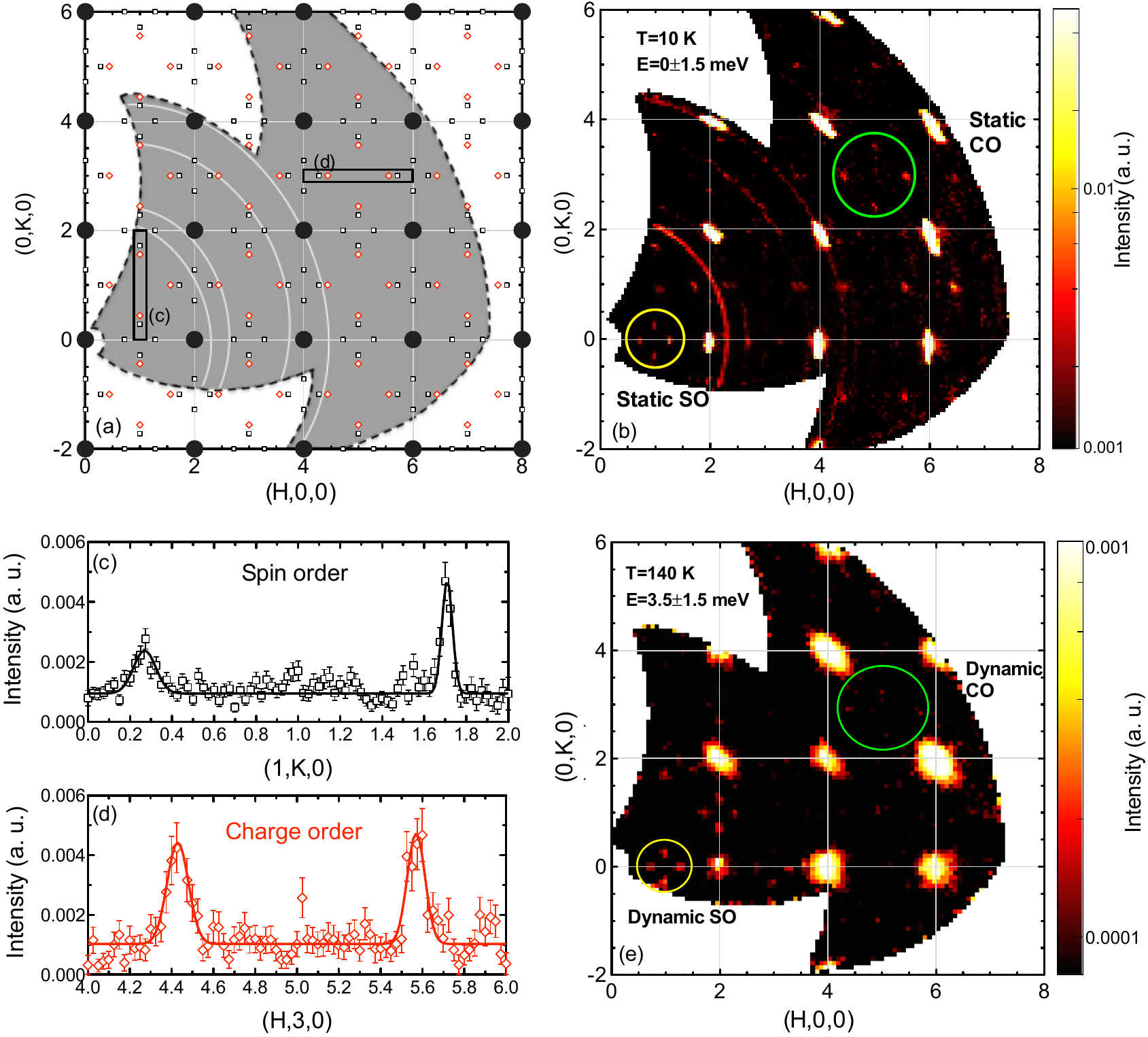}
    \caption{\label{fig:Fig1} (color online). (a) Schematic diagram indicating relative locations of  fundamental Bragg peaks (black solid circles), charge-order peaks (red hollow diamonds) and spin-order peaks (black hollow squares) in the $(H, K, 0)$ plane of \LSNOn.  Shaded area illustrates the scanning range where data have been collected at 10~K.  White arcs represent aluminum powder rings. (b) Constant energy slice of $(H, K, 0)$ plane at 10~K. Elastic scattering intensities have been integrated with $-0.2\leq L \leq0.2$, $-1.5\leq E \leq1.5$~meV. (c) Scans through magnetic peaks along $\mathbf{Q} = (1,K,0)$. (d) Scans through charge-order peaks along $\mathbf{Q} = (H, 3, 0)$. (e) Inelastic neutron scattering (integrated over $2\leq E \leq 5$ meV, $-0.2\leq L \leq0.2$) measured at $T_{\text{so}}=140$~K. Bright spots centered at fundamental Bragg wave vectors represent acoustic phonons. Weak spots circled in yellow (green) represent the dynamic spin (charge) stripes at small (large) \textbf{Q}.}
\end{figure*}

The charge and spin stripes that develop in LSNO run diagonally with respect to the square lattice of Ni atoms in the NiO$_2$ planes.  While the average crystal structure is tetragonal (space group $I4/mmm$) in the relevant doping range \cite{huck04b}, it is easier to characterize the stripe wave vectors if we use a unit cell of doubled volume (space group $F4/mmm$); for $x=0.25$, this corresponds to lattice parameters $a=b=5.42$~\AA\ and $c=12.64$~\AA.    With this choice, the charge and spin wave vectors are
\begin{equation}
\mathbf{ q_{co}} = (2\epsilon, 0, 1),
\qquad
\mathbf{ q_{so}} = (1\pm\epsilon, 0, 0),
\end{equation}
where the coordinates are in reciprocal lattice units $(2\pi/a,2\pi/a,2\pi/c)$; there is also a stripe twin domain rotated by $90^\circ$ in the NiO$_2$ plane.  For the fundamental Bragg peaks, ${\bf G}=(H,K,L)$ the indices must be all even or all odd.  It follows that the allowed superlattice peaks in the $(H,K,0)$ reciprocal plane are
\begin{align}
\mathbf{ G'\pm q}_{\rm co} & = (2m+1\pm 2\epsilon, 2n+1, 0),\\
\mathbf{ G\pm q}_{\rm so} & = (2m+1\pm\epsilon, 2n, 0),
\end{align}
with $m,n = $ integers.  These positions are illustrated in Fig.~\ref{fig:Fig1}(a).  For the case of $x=0.33$, where $\epsilon=0.33$ \cite{lee97}, the charge and spin peaks overlap, making it difficult to establish the relative contributions to each peak.  For this reason, we have chosen to focus on $x=0.25$ (with $\epsilon=0.28$ \cite{anis14}) where the charge and spin peaks are distinct.

Neutron scattering measurements were carried out on the time-of-flight Hybrid Spectrometer (HYSPEC) at BL-14B of Spallation Neutron Source, Oak Ridge National Laboratory \cite{hyspec15}. The \LSNOn\ single-crystal sample, with a mass of $\sim10$~g, was grown by the traveling-solvent floating zone method at Brookhaven and was characterized previously \cite{anis14}.  For this experiment, it
was mounted in a Displex closed-cycle cryostat with the $(HK0)$ plane horizontal and $c$ axis vertical, perpendicular to the incident neutron beam.  The incident energy was 
50~meV.  Data analysis was performed with the MANTID \cite{mantid14} and DAVE \cite{dave09} software packages.  For further details, see \cite{suppl}.

\begin{figure}[h]
 \centering
    \includegraphics[width=0.9\columnwidth]{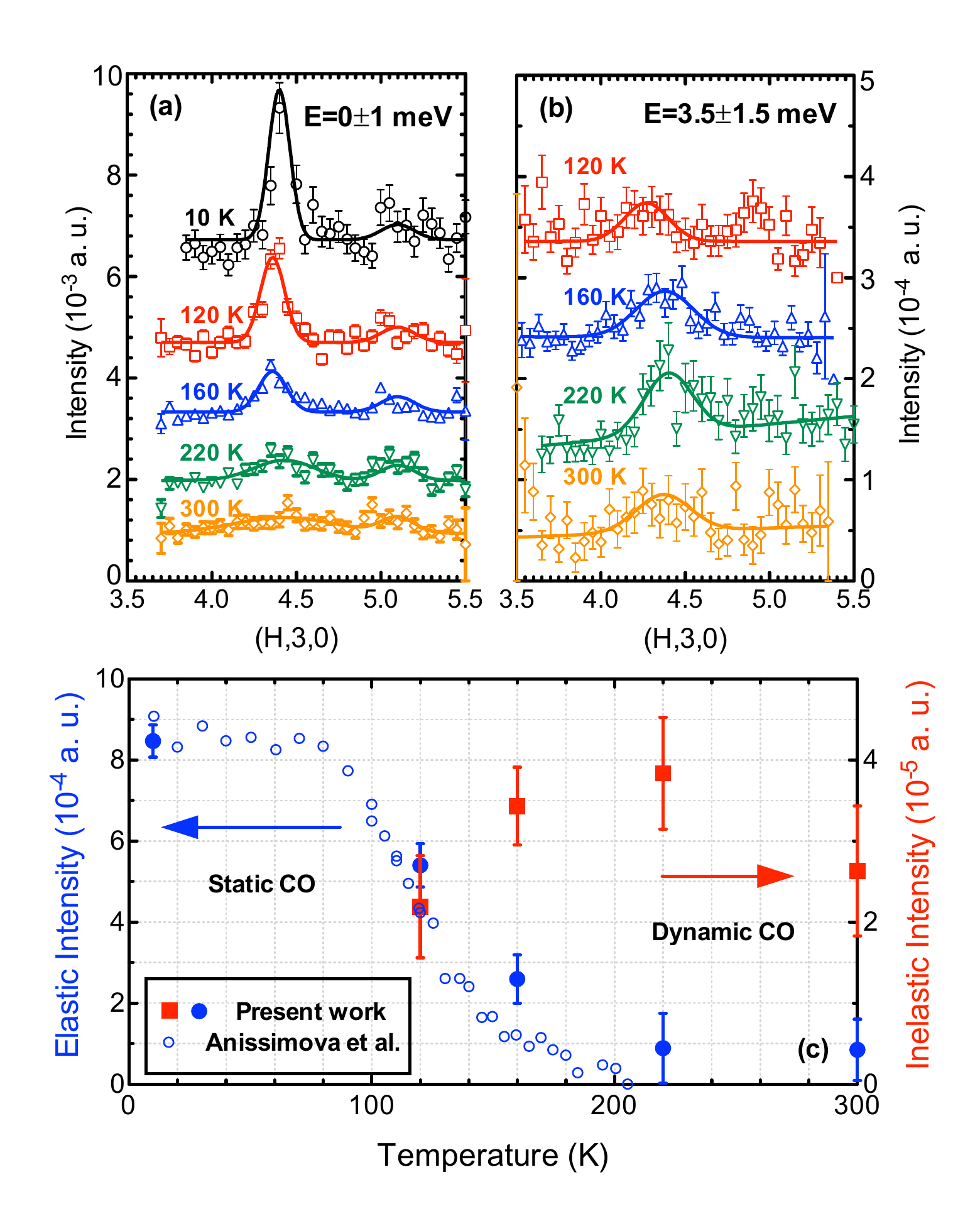}
    \caption{\label{fig:Fig2} (color online). (a) Elastic scattering ($-0.2 \leq L \leq 0.2$, $-1\leq E \leq1$~meV) associated with static charge order along $(H, 3, 0)$ measured at different temperatures. The temperature-independent peaks around $(5.1, 3, 0)$ are powder diffraction from the aluminum sample holder. (b) Inelastic signal from dynamical charge order ($-0.2 \leq L \leq 0.2$, $2\leq E \leq 5$~meV) measured at different temperatures. In both (a) and (b), intensities are normalized to incident flux. Data sets have been shifted for clarity; solid lines represent Gaussian fits. (c) Temperature dependence of the integrated intensity of static (blue filled circles) and dynamic (red filled squares) charge-stripe correlations. For comparison, triple-axis results of static charge order (blue open circles) from Anissimova et al.\cite{anis14} are included. }
\end{figure}

Characterizations of the spin and charge scattering are summarized in Fig.~\ref{fig:Fig1}.  The magnetic peaks should be strong at small $Q$ and decrease in intensity at large $Q$ due to the fall of the magnetic form factor \cite{anis14}.  In contrast, the charge order scattering is detected through associated atomic displacements, for which the intensity should grow roughly as $Q^2$.  In the map of low-temperature elastic scattering shown in Fig.~\ref{fig:Fig1}(b), we expect that the peaks within the small and large circles should correspond to spin and charge order, respectively.  This is confirmed by looking at particular line cuts: Fig.~\ref{fig:Fig1}(c) shows spin order peaks at $(1,\epsilon,0)$ and $(1,2-\epsilon,0)$ with no significant weight at the charge order positions $(1,1\pm2\epsilon,0)$, while Fig.~\ref{fig:Fig1}(c) shows only charge order peaks at $(5\pm2\epsilon,3,0)$; Gaussian peak fitting gives $\epsilon=0.28$.  From the map of low-energy inelastic scattering obtained at 140~K, shown in Fig.~\ref{fig:Fig1}(e), we see that the spin fluctuation scattering also falls off with $Q$, as expected, so that the incommensurate peaks in the large circle must correspond to charge stripe fluctuations.  From here on, we will focus on the scattering near ${\bf Q}_{\rm co}^*=(4.44, 3, 0)$.

\begin{figure}[ht]
 \centering
    \includegraphics[width=0.9\columnwidth]{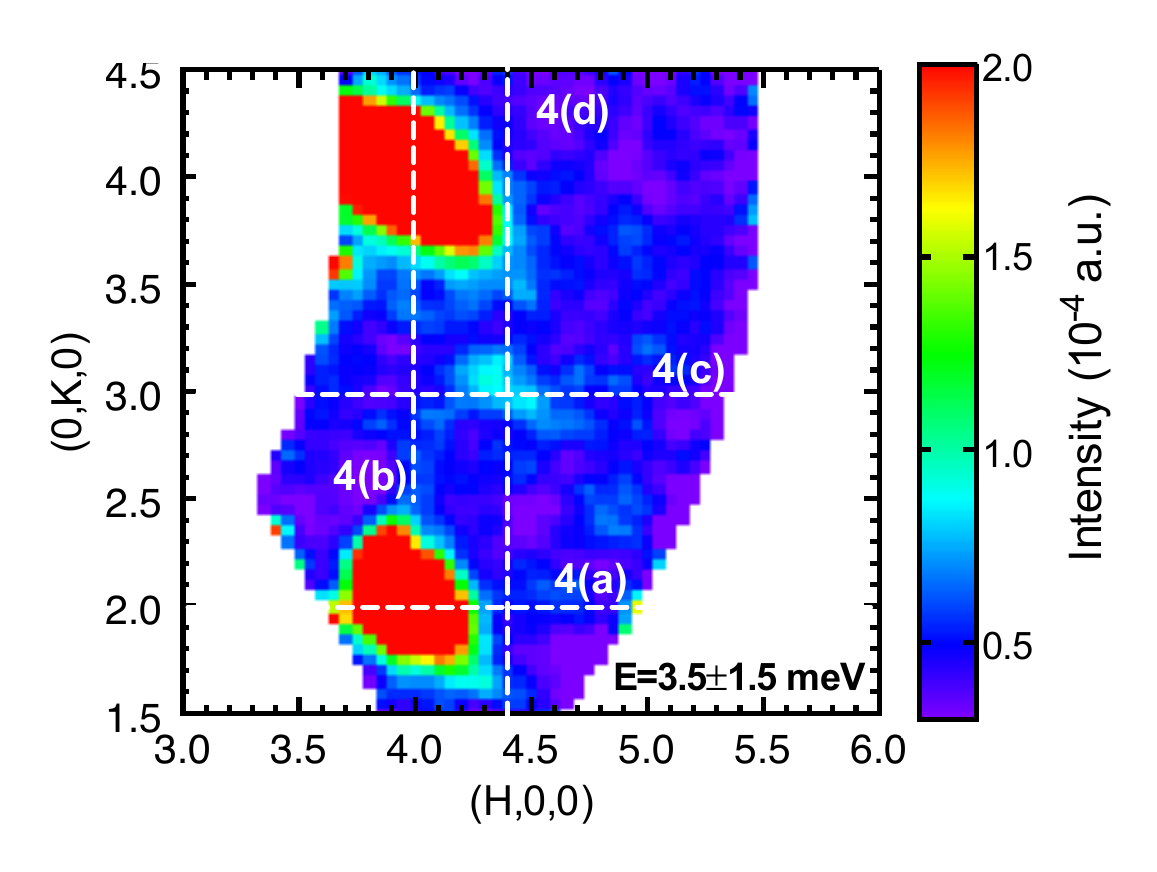}
    \caption{\label{fig:Fig3} (color online). Constant energy slice ($2\leq E \leq5$ meV, $-0.2\leq L\leq0.2$) around the charge-stripe peak at ${\bf Q}^*_{\rm co}=(4.44, 3, 0)$ at 160~K, plotted in the $(HK0)$ plane. The strongest scattering, centered at ${\bf G}=(4, 4, 0)$ and $(2, 4, 0)$, comes from acoustic phonons. White dashed lines and letters indicate the direction of corresponding slices in Fig.~\hyperref[fig:Fig4]{4}. }
\end{figure}

\begin{figure*}[ht]
 \centering
    \includegraphics[width=1.8\columnwidth]{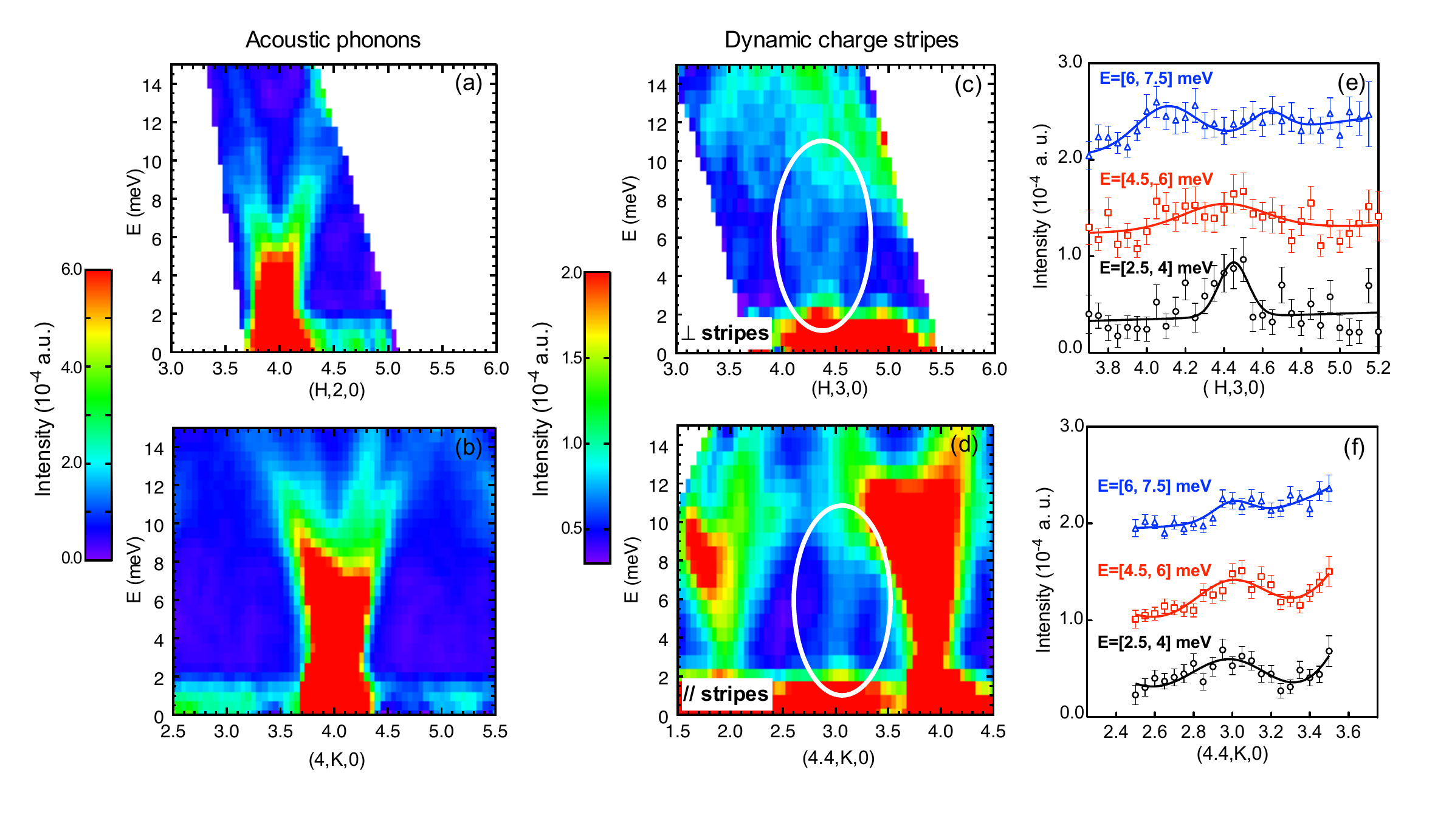}
    \caption{\label{fig:Fig4} (color online). Low-energy excitations of lattice and charge stripes in \LSNOn\ at 160~K. (a) Acoustic phonons dispersing from the Bragg peak ${\bf Q}=(4, 2, 0)$ along [100],  integrated over $1.9\leq K\leq2.1$, and (b) from ${\bf Q}=(4, 4, 0)$ along [010], integrated over $3.75\leq H\leq4.25$. (c,d) Scattered intensity as a function of $E$ vs \textbf{Q} through ${\bf Q}^*_{\rm co}$ along [100], integrated over $2.9\leq K\leq3.1$ (c), and along [010], integrated over $4.1\leq H\leq4.7$ (d). Charge-stripe fluctuations are indicated by white ovals. (e)  Cuts through the charge stripe fluctuations in (c), integrated over fixed energy bands, as noted in the labels;  (f) cuts through the charge stripe fluctuations in (d). In all cases, data have been integrated over $-0.2\leq L\leq0.2$.  In (a)-(d), a {\bf Q}-independent incoherent elastic scattering contribution, broadened by instrumental energy resolution, has been subtracted.  Lines through the data points in (e) and (d) are fitted Gaussian peaks plus background; in (e), the background is linear in $H$, while in (f), the background is a constant plus the tail of a Gaussian peak at large $K$ to account for the neighboring acoustic phonons.}
\end{figure*}

The temperature dependence of the elastic and low-energy inelastic scattering at ${\bf Q}_{\rm co}^*$ are presented in the form of line cuts in Fig.~\ref{fig:Fig2}(a) and (b).  From the fitted Gaussian peaks, indicated by the solid lines, we obtain the integrated elastic and inelastic intensities that are plotted vs.\ temperature in Fig.~\ref{fig:Fig2}(c); for comparison, the previous triple-axis measurements of the elastic charge-order intensity are included \cite{anis14}.  While it is established that the charge modulation is unidirectional \cite{li03,huck06}, the presence of finite correlation lengths for charge order within and between planes \cite{tran96a,yosh00} means that we have, at best, a stripe (or smectic) glass, due to the quenched disorder associated with the Sr dopant ions.  The elastic charge-order scattering decreases rapidly as the spin order disappears at $T_{\rm so}\approx140$~K.  Previous studies have shown that, above this point, correlation lengths shrink \cite{tran96a,lee02,spen05,lee12} and finite optical conductivity turns on \cite{home03,lloy08,cosl13}, indicating the importance of fluctuations and the absence of long-range order.  Consistent with this, the charge-stripe fluctuations detected by inelastic scattering only become significant above $T_{\rm so}$.  (The small but finite ``elastic'' signal above 210~K appears to come from integration over quasi-elastic scattering.)  {\newr While any semblance of proper smectic order is certainly destroyed above $T_{\rm so}$ \cite{zach03}, there remains the possibility of vestigial nematic order \cite{nie14}.}

In an actual liquid-crystal system, one would distinguish the nematic from the smectic phase by the development of anisotropic peak widths \cite{[][{, Sec.~2.7.}]chai95}.  Here, the finite width due to disorder prevents that; instead, we look for anisotropy in the dispersion of the charge stripe fluctuations.  Of course, the crystal's symmetry is tetragonal, and without a symmetry-breaking field, we can never observe long-range nematic order; nevertheless, by measuring at a wave vector corresponding to charge-density modulations, we can selectively look at domains with the same modulation orientation.  If we can resolve an anisotropic dispersion, then we conclude that the system has spontaneously broken the local rotational symmetry.

To obtain meaningful data, it is necessary to measure at locations and temperatures where the signal is significant.  Figure~\ref{fig:Fig3} shows a reciprocal-space map of low-energy excitations in the vicinity of ${\bf Q}_{\rm co}^*$ measured at 160~K, where the fluctuation intensity initially rises to a reasonable level.  The dashed lines denote the paths of the energy vs.\ {\bf Q} slices plotted in Fig.~\ref{fig:Fig4}.   Measurements at 160~K along and transverse to the modulation direction are plotted in Fig.~\ref{fig:Fig4}(c) and (d), respectively;  
in each case, the stripe fluctuations are circled, and in \ref{fig:Fig4}(d) 
they sit between acoustic phonons dispersing from neighboring Bragg peaks (see Fig.~\ref{fig:Fig3}).  Figures~4(e) and (f) show constant-energy cuts through the dispersion, integrated over 1.5-meV bands.  (For similar results at 220~K, see \cite{suppl}.) The data in Fig.~\ref{fig:Fig4}(a) and (b) show acoustic phonons at fundamental Bragg reflections for reference.  Note that the stripe fluctuations have an intensity above background that is about two orders of magnitude weaker than that of the acoustic modes (whose true intensity is masked by saturation).  

The window for viewing the stripe fluctuations is small---by an energy of 9 meV or so they 
run into the transverse acoustic modes dispersing from the $(4,4,0)$ Bragg peak (see Fig.~\ref{fig:Fig3}).   It is notable that we do not see any significant interaction between the stripe fluctuations and the acoustic mode.  In contrast to the soft-mode behavior detected in association with the charge-density-wave order in underdoped YBa$_2$Cu$_3$O$_{6+x}$ \cite{leta14,blac13b}, we appear to have overlapping dispersions.  The slowest dispersion is in the direction perpendicular to the stripes, where the effective velocity is comparable to that of transverse acoustic modes shown in Fig.~\ref{fig:Fig4}(b).  As this is the modulation direction, it is also associated with the observation that the incommensurability $\epsilon$ increases towards $1/3$ as the temperature approaches $T_{\rm co}$ \cite{tran96a,ishi04}.
The velocity parallel to the stripes is not resolved but might be comparable to that of the longitudinal acoustic mode that is resolved at energies above 10 meV in Fig.~\ref{fig:Fig4}(a) and (b).  Note, however, that the comparison with phonon velocities is only to provide relevant scale.  The specific anisotropy of the fluctuations is not consistent with what one observes for normal acoustic phonons about a structural superlattice peak.

The combination of the characteristic modulation wave vector and the dispersion anisotropy provide a strong case for the presence of electronic nematic order.  This is of particular significance given the four-fold rotational symmetry of the average crystal structure.  (Note that some of the best evidence for nematic order in cuprates occurs in samples where the crystal structure has reduced rotational symmetry \cite{daou10}.)  The nematic order in this case is formed of fluctuating charge stripes, as in the original proposal by Kivelson, Fradkin, and Emery \cite{kive98}.  The response at 160 K is compatible with nematic order as a vestigial version of static charge density waves \cite{nie14} similar to what has been observed by scanning tunneling microscopy in \bscco\ \cite{lawl10,mesa11}.  The charge stripe correlations certainly involve a coupling to the lattice, as without nuclear displacements we would not be able to detect the charge stripes with neutrons.  In a sense, this is polaronic; however, it is not made up of individually dressed holes, but is, instead, an emergent collective state, one not easily captured by current {\it ab initio} techniques.  We hope that these results will inspire new theoretical efforts and further experimental developments.

We thank M.~K.~Graves-Brook for valuable assistance at HYSPEC and A.~T.~Savici for help with the data analysis using Mantid.  We are grateful to L. M. Nie for helpful comments.  Work at Brookhaven is supported by the Office of Basic Energy Sciences (BES), Materials Sciences and Engineering Division (MSED), U.S. Department of Energy (DOE) under Contract No.\ DE-SC0012704. D.R. is also supported by BES/MSED under Contract No.\ DE-SC0006939.  A portion of this research used resources at the Spallation Neutron Source, a DOE Office of Science User Facility operated by the Oak Ridge National Laboratory. 

\bibliography{LNO,theory,neutrons,misc}

\begin{thebibliography}{51}%
\makeatletter
\providecommand \@ifxundefined [1]{%
 \@ifx{#1\undefined}
}%
\providecommand \@ifnum [1]{%
 \ifnum #1\expandafter \@firstoftwo
 \else \expandafter \@secondoftwo
 \fi
}%
\providecommand \@ifx [1]{%
 \ifx #1\expandafter \@firstoftwo
 \else \expandafter \@secondoftwo
 \fi
}%
\providecommand \natexlab [1]{#1}%
\providecommand \enquote  [1]{``#1''}%
\providecommand \bibnamefont  [1]{#1}%
\providecommand \bibfnamefont [1]{#1}%
\providecommand \citenamefont [1]{#1}%
\providecommand \href@noop [0]{\@secondoftwo}%
\providecommand \href [0]{\begingroup \@sanitize@url \@href}%
\providecommand \@href[1]{\@@startlink{#1}\@@href}%
\providecommand \@@href[1]{\endgroup#1\@@endlink}%
\providecommand \@sanitize@url [0]{\catcode `\\12\catcode `\$12\catcode
  `\&12\catcode `\#12\catcode `\^12\catcode `\_12\catcode `\%12\relax}%
\providecommand \@@startlink[1]{}%
\providecommand \@@endlink[0]{}%
\providecommand \url  [0]{\begingroup\@sanitize@url \@url }%
\providecommand \@url [1]{\endgroup\@href {#1}{\urlprefix }}%
\providecommand \urlprefix  [0]{URL }%
\providecommand \Eprint [0]{\href }%
\providecommand \doibase [0]{http://dx.doi.org/}%
\providecommand \selectlanguage [0]{\@gobble}%
\providecommand \bibinfo  [0]{\@secondoftwo}%
\providecommand \bibfield  [0]{\@secondoftwo}%
\providecommand \translation [1]{[#1]}%
\providecommand \BibitemOpen [0]{}%
\providecommand \bibitemStop [0]{}%
\providecommand \bibitemNoStop [0]{.\EOS\space}%
\providecommand \EOS [0]{\spacefactor3000\relax}%
\providecommand \BibitemShut  [1]{\csname bibitem#1\endcsname}%
\let\auto@bib@innerbib\@empty
\bibitem [{\citenamefont {Tranquada}\ \emph {et~al.}(1994)\citenamefont
  {Tranquada}, \citenamefont {Buttrey}, \citenamefont {Sachan},\ and\
  \citenamefont {Lorenzo}}]{tran94a}%
  \BibitemOpen
  \bibfield  {author} {\bibinfo {author} {\bibfnamefont {J.~M.}\ \bibnamefont
  {Tranquada}}, \bibinfo {author} {\bibfnamefont {D.~J.}\ \bibnamefont
  {Buttrey}}, \bibinfo {author} {\bibfnamefont {V.}~\bibnamefont {Sachan}}, \
  and\ \bibinfo {author} {\bibfnamefont {J.~E.}\ \bibnamefont {Lorenzo}},\
  }\bibfield  {title} {\enquote {\bibinfo {title} {{Simultaneous Ordering of
  Holes and Spins in ${\mathrm{La}}_{2}$Ni${\mathrm{O}}_{4.125}$}},}\
  }\href@noop {} {\bibfield  {journal} {\bibinfo  {journal} {Phys. Rev. Lett.}\
  }\textbf {\bibinfo {volume} {73}},\ \bibinfo {pages} {1003--1006} (\bibinfo
  {year} {1994})}\BibitemShut {NoStop}%
\bibitem [{\citenamefont {Lee}\ and\ \citenamefont {Cheong}(1997)}]{lee97}%
  \BibitemOpen
  \bibfield  {author} {\bibinfo {author} {\bibfnamefont {S.-H.}\ \bibnamefont
  {Lee}}\ and\ \bibinfo {author} {\bibfnamefont {S-W.}\ \bibnamefont
  {Cheong}},\ }\bibfield  {title} {\enquote {\bibinfo {title} {{Melting of
  Quasi-Two-Dimensional Charge Stripes in
  ${\mathrm{La}}_{5\mathrm{/}3}{\mathrm{Sr}}_{1\mathrm{/}3}{\mathrm{NiO}}_{4}$}},}\
  }\href@noop {} {\bibfield  {journal} {\bibinfo  {journal} {Phys. Rev. Lett.}\
  }\textbf {\bibinfo {volume} {79}},\ \bibinfo {pages} {2514--2517} (\bibinfo
  {year} {1997})}\BibitemShut {NoStop}%
\bibitem [{\citenamefont {Yoshizawa}\ \emph {et~al.}(2000)\citenamefont
  {Yoshizawa}, \citenamefont {Kakeshita}, \citenamefont {Kajimoto},
  \citenamefont {Tanabe}, \citenamefont {Katsufuji},\ and\ \citenamefont
  {Tokura}}]{yosh00}%
  \BibitemOpen
  \bibfield  {author} {\bibinfo {author} {\bibfnamefont {H.}~\bibnamefont
  {Yoshizawa}}, \bibinfo {author} {\bibfnamefont {T.}~\bibnamefont
  {Kakeshita}}, \bibinfo {author} {\bibfnamefont {R.}~\bibnamefont {Kajimoto}},
  \bibinfo {author} {\bibfnamefont {T.}~\bibnamefont {Tanabe}}, \bibinfo
  {author} {\bibfnamefont {T.}~\bibnamefont {Katsufuji}}, \ and\ \bibinfo
  {author} {\bibfnamefont {Y.}~\bibnamefont {Tokura}},\ }\bibfield  {title}
  {\enquote {\bibinfo {title} {{Stripe order at low temperatures in
  La$_{2-x}$Sr$_{x}$NiO$_{4}$ with $0.289\lesssim x\lesssim0.5$}},}\
  }\href@noop {} {\bibfield  {journal} {\bibinfo  {journal} {Phys. Rev. B}\
  }\textbf {\bibinfo {volume} {61}},\ \bibinfo {pages} {R854--R857} (\bibinfo
  {year} {2000})}\BibitemShut {NoStop}%
\bibitem [{\citenamefont {Ulbrich}\ and\ \citenamefont
  {Braden}(2012)}]{ulbr12b}%
  \BibitemOpen
  \bibfield  {author} {\bibinfo {author} {\bibfnamefont {Holger}\ \bibnamefont
  {Ulbrich}}\ and\ \bibinfo {author} {\bibfnamefont {Markus}\ \bibnamefont
  {Braden}},\ }\bibfield  {title} {\enquote {\bibinfo {title} {{Neutron
  scattering studies on stripe phases in non-cuprate materials}},}\ }\href@noop
  {} {\bibfield  {journal} {\bibinfo  {journal} {Physica C}\ }\textbf {\bibinfo
  {volume} {481}},\ \bibinfo {pages} {31--45} (\bibinfo {year}
  {2012})}\BibitemShut {NoStop}%
\bibitem [{\citenamefont {Keimer}\ \emph {et~al.}(2015)\citenamefont {Keimer},
  \citenamefont {Kivelson}, \citenamefont {Norman}, \citenamefont {Uchida},\
  and\ \citenamefont {Zaanen}}]{keim15}%
  \BibitemOpen
  \bibfield  {author} {\bibinfo {author} {\bibfnamefont {B.}~\bibnamefont
  {Keimer}}, \bibinfo {author} {\bibfnamefont {S.~A.}\ \bibnamefont
  {Kivelson}}, \bibinfo {author} {\bibfnamefont {M.~R.}\ \bibnamefont
  {Norman}}, \bibinfo {author} {\bibfnamefont {S.}~\bibnamefont {Uchida}}, \
  and\ \bibinfo {author} {\bibfnamefont {J.}~\bibnamefont {Zaanen}},\
  }\bibfield  {title} {\enquote {\bibinfo {title} {{From quantum matter to
  high-temperature superconductivity in copper oxides}},}\ }\href@noop {}
  {\bibfield  {journal} {\bibinfo  {journal} {Nature}\ }\textbf {\bibinfo
  {volume} {518}},\ \bibinfo {pages} {179--186} (\bibinfo {year}
  {2015})}\BibitemShut {NoStop}%
\bibitem [{\citenamefont {Comin}\ and\ \citenamefont
  {Damascelli}(2016)}]{comi16}%
  \BibitemOpen
  \bibfield  {author} {\bibinfo {author} {\bibfnamefont {Riccardo}\
  \bibnamefont {Comin}}\ and\ \bibinfo {author} {\bibfnamefont {Andrea}\
  \bibnamefont {Damascelli}},\ }\bibfield  {title} {\enquote {\bibinfo {title}
  {{Resonant X-Ray Scattering Studies of Charge Order in Cuprates}},}\ }\href
  {\doibase 10.1146/annurev-conmatphys-031115-011401} {\bibfield  {journal}
  {\bibinfo  {journal} {Annu. Rev. Condens. Matter Phys.}\ }\textbf {\bibinfo
  {volume} {7}},\ \bibinfo {pages} {369--405} (\bibinfo {year}
  {2016})}\BibitemShut {NoStop}%
\bibitem [{\citenamefont {Brom}\ and\ \citenamefont {Zaanen}(2003)}]{brom03}%
  \BibitemOpen
  \bibfield  {author} {\bibinfo {author} {\bibfnamefont {Hans~B.}\ \bibnamefont
  {Brom}}\ and\ \bibinfo {author} {\bibfnamefont {Jan}\ \bibnamefont
  {Zaanen}},\ }\bibfield  {title} {\enquote {\bibinfo {title} {{Magnetic
  ordering phenomena and dynamic fluctuations in cuprate superconductors and
  insulating nickelates}},}\ }in\ \href {\doibase
  http://dx.doi.org/10.1016/S1567-2719(03)15004-4} {\emph {\bibinfo {booktitle}
  {Handbook of Magnetic Materials}}},\ \bibinfo {series} {Handbook of Magnetic
  Materials}, Vol.~\bibinfo {volume} {15}\ (\bibinfo  {publisher} {Elsevier},\
  \bibinfo {year} {2003})\ pp.\ \bibinfo {pages} {379--496}\BibitemShut
  {NoStop}%
\bibitem [{\citenamefont {Tranquada}(2013)}]{tran13a}%
  \BibitemOpen
  \bibfield  {author} {\bibinfo {author} {\bibfnamefont {John~M.}\ \bibnamefont
  {Tranquada}},\ }\bibfield  {title} {\enquote {\bibinfo {title} {{Spins,
  stripes, and superconductivity in hole-doped cuprates}},}\ }\href@noop {}
  {\bibfield  {journal} {\bibinfo  {journal} {AIP Conf. Proc.}\ }\textbf
  {\bibinfo {volume} {1550}},\ \bibinfo {pages} {114--187} (\bibinfo {year}
  {2013})}\BibitemShut {NoStop}%
\bibitem [{\citenamefont {Zaanen}\ and\ \citenamefont
  {Littlewood}(1994)}]{zaan94}%
  \BibitemOpen
  \bibfield  {author} {\bibinfo {author} {\bibfnamefont {J.}~\bibnamefont
  {Zaanen}}\ and\ \bibinfo {author} {\bibfnamefont {P.~B.}\ \bibnamefont
  {Littlewood}},\ }\bibfield  {title} {\enquote {\bibinfo {title} {{Freezing
  electronic correlations by polaronic instabilities in doped
  ${\mathrm{La}}_{2}$${\mathrm{NiO}}_{4}$}},}\ }\href@noop {} {\bibfield
  {journal} {\bibinfo  {journal} {Phys. Rev. B}\ }\textbf {\bibinfo {volume}
  {50}},\ \bibinfo {pages} {7222--7225} (\bibinfo {year} {1994})}\BibitemShut
  {NoStop}%
\bibitem [{\citenamefont {Hotta}\ and\ \citenamefont {Dagotto}(2004)}]{hott04}%
  \BibitemOpen
  \bibfield  {author} {\bibinfo {author} {\bibfnamefont {Takashi}\ \bibnamefont
  {Hotta}}\ and\ \bibinfo {author} {\bibfnamefont {Elbio}\ \bibnamefont
  {Dagotto}},\ }\bibfield  {title} {\enquote {\bibinfo {title} {{Orbital
  Ordering, New Phases, and Stripe Formation in Doped Layered Nickelates}},}\
  }\href {\doibase 10.1103/PhysRevLett.92.227201} {\bibfield  {journal}
  {\bibinfo  {journal} {Phys. Rev. Lett.}\ }\textbf {\bibinfo {volume} {92}},\
  \bibinfo {pages} {227201} (\bibinfo {year} {2004})}\BibitemShut {NoStop}%
\bibitem [{\citenamefont {Raczkowski}\ \emph {et~al.}(2006)\citenamefont
  {Raczkowski}, \citenamefont {Fr\'esard},\ and\ \citenamefont
  {Ole\ifmmode~\acute{s}\else \'{s}\fi{}}}]{racz06}%
  \BibitemOpen
  \bibfield  {author} {\bibinfo {author} {\bibfnamefont {Marcin}\ \bibnamefont
  {Raczkowski}}, \bibinfo {author} {\bibfnamefont {Raymond}\ \bibnamefont
  {Fr\'esard}}, \ and\ \bibinfo {author} {\bibfnamefont {Andrzej~M.}\
  \bibnamefont {Ole\ifmmode~\acute{s}\else \'{s}\fi{}}},\ }\bibfield  {title}
  {\enquote {\bibinfo {title} {{Microscopic origin of diagonal stripe phases in
  doped nickelates}},}\ }\href {\doibase 10.1103/PhysRevB.73.094429} {\bibfield
   {journal} {\bibinfo  {journal} {Phys. Rev. B}\ }\textbf {\bibinfo {volume}
  {73}},\ \bibinfo {pages} {094429} (\bibinfo {year} {2006})}\BibitemShut
  {NoStop}%
\bibitem [{\citenamefont {Yamamoto}\ \emph {et~al.}(2007)\citenamefont
  {Yamamoto}, \citenamefont {Fujiwara},\ and\ \citenamefont
  {Hatsugai}}]{yama07}%
  \BibitemOpen
  \bibfield  {author} {\bibinfo {author} {\bibfnamefont {Susumu}\ \bibnamefont
  {Yamamoto}}, \bibinfo {author} {\bibfnamefont {Takeo}\ \bibnamefont
  {Fujiwara}}, \ and\ \bibinfo {author} {\bibfnamefont {Yasuhiro}\ \bibnamefont
  {Hatsugai}},\ }\bibfield  {title} {\enquote {\bibinfo {title} {{Electronic
  structure of charge and spin stripe order in
  ${\mathrm{La}}_{2\ensuremath{-}x}{\mathrm{Sr}}_{x}\mathrm{Ni}{\mathrm{O}}_{4}$
  $(x=\frac{1}{3},\frac{1}{2})$}},}\ }\href {\doibase
  10.1103/PhysRevB.76.165114} {\bibfield  {journal} {\bibinfo  {journal} {Phys.
  Rev. B}\ }\textbf {\bibinfo {volume} {76}},\ \bibinfo {pages} {165114}
  (\bibinfo {year} {2007})}\BibitemShut {NoStop}%
\bibitem [{\citenamefont {Schwingenschl\"ogl}\ \emph
  {et~al.}(2008)\citenamefont {Schwingenschl\"ogl}, \citenamefont {Schuster},\
  and\ \citenamefont {Fr\'esard}}]{schw08}%
  \BibitemOpen
  \bibfield  {author} {\bibinfo {author} {\bibfnamefont {U.}~\bibnamefont
  {Schwingenschl\"ogl}}, \bibinfo {author} {\bibfnamefont {C.}~\bibnamefont
  {Schuster}}, \ and\ \bibinfo {author} {\bibfnamefont {R.}~\bibnamefont
  {Fr\'esard}},\ }\bibfield  {title} {\enquote {\bibinfo {title} {{Magnetic
  ordering in the striped nickelate La$_{5/3}$Sr$_{1/3}$NiO$_4$: A band
  structure point of view}},}\ }\href
  {http://stacks.iop.org/0295-5075/81/i=2/a=27002} {\bibfield  {journal}
  {\bibinfo  {journal} {Europhys. Lett.}\ }\textbf {\bibinfo {volume} {81}},\
  \bibinfo {pages} {27002} (\bibinfo {year} {2008})}\BibitemShut {NoStop}%
\bibitem [{\citenamefont {Katsufuji}\ \emph {et~al.}(1996)\citenamefont
  {Katsufuji}, \citenamefont {Tanabe}, \citenamefont {Ishikawa}, \citenamefont
  {Fukuda}, \citenamefont {Arima},\ and\ \citenamefont {Tokura}}]{kats96}%
  \BibitemOpen
  \bibfield  {author} {\bibinfo {author} {\bibfnamefont {T.}~\bibnamefont
  {Katsufuji}}, \bibinfo {author} {\bibfnamefont {T.}~\bibnamefont {Tanabe}},
  \bibinfo {author} {\bibfnamefont {T.}~\bibnamefont {Ishikawa}}, \bibinfo
  {author} {\bibfnamefont {Y.}~\bibnamefont {Fukuda}}, \bibinfo {author}
  {\bibfnamefont {T.}~\bibnamefont {Arima}}, \ and\ \bibinfo {author}
  {\bibfnamefont {Y.}~\bibnamefont {Tokura}},\ }\bibfield  {title} {\enquote
  {\bibinfo {title} {{Optical spectroscopy of the charge-ordering transition in
  ${\mathrm{La}}_{1.67}$${\mathrm{Sr}}_{0.33}$${\mathrm{NiO}}_{4}$}},}\
  }\href@noop {} {\bibfield  {journal} {\bibinfo  {journal} {Phys. Rev. B}\
  }\textbf {\bibinfo {volume} {54}},\ \bibinfo {pages} {R14230--R14233}
  (\bibinfo {year} {1996})}\BibitemShut {NoStop}%
\bibitem [{\citenamefont {Homes}\ \emph {et~al.}(2003)\citenamefont {Homes},
  \citenamefont {Tranquada}, \citenamefont {Li}, \citenamefont {Moodenbaugh},\
  and\ \citenamefont {Buttrey}}]{home03}%
  \BibitemOpen
  \bibfield  {author} {\bibinfo {author} {\bibfnamefont {C.~C.}\ \bibnamefont
  {Homes}}, \bibinfo {author} {\bibfnamefont {J.~M.}\ \bibnamefont
  {Tranquada}}, \bibinfo {author} {\bibfnamefont {Q.}~\bibnamefont {Li}},
  \bibinfo {author} {\bibfnamefont {A.~R.}\ \bibnamefont {Moodenbaugh}}, \ and\
  \bibinfo {author} {\bibfnamefont {D.~J.}\ \bibnamefont {Buttrey}},\
  }\bibfield  {title} {\enquote {\bibinfo {title} {{Mid-infrared conductivity
  from mid-gap states associated with charge stripes}},}\ }\href@noop {}
  {\bibfield  {journal} {\bibinfo  {journal} {Phys. Rev. B}\ }\textbf {\bibinfo
  {volume} {67}},\ \bibinfo {pages} {184516} (\bibinfo {year}
  {2003})}\BibitemShut {NoStop}%
\bibitem [{\citenamefont {Coslovich}\ \emph {et~al.}(2013)\citenamefont
  {Coslovich}, \citenamefont {Huber}, \citenamefont {Lee}, \citenamefont
  {Chuang}, \citenamefont {Zhu}, \citenamefont {Sasagawa}, \citenamefont
  {Hussain}, \citenamefont {Bechtel}, \citenamefont {Martin}, \citenamefont
  {Shen}, \citenamefont {Schoenlein},\ and\ \citenamefont {Kaindl}}]{cosl13}%
  \BibitemOpen
  \bibfield  {author} {\bibinfo {author} {\bibfnamefont {G.}~\bibnamefont
  {Coslovich}}, \bibinfo {author} {\bibfnamefont {B.}~\bibnamefont {Huber}},
  \bibinfo {author} {\bibfnamefont {W.~S.}\ \bibnamefont {Lee}}, \bibinfo
  {author} {\bibfnamefont {Y.~D.}\ \bibnamefont {Chuang}}, \bibinfo {author}
  {\bibfnamefont {Y.}~\bibnamefont {Zhu}}, \bibinfo {author} {\bibfnamefont
  {T.}~\bibnamefont {Sasagawa}}, \bibinfo {author} {\bibfnamefont
  {Z.}~\bibnamefont {Hussain}}, \bibinfo {author} {\bibfnamefont {H.~A.}\
  \bibnamefont {Bechtel}}, \bibinfo {author} {\bibfnamefont {M.~C.}\
  \bibnamefont {Martin}}, \bibinfo {author} {\bibfnamefont {Z.~X.}\
  \bibnamefont {Shen}}, \bibinfo {author} {\bibfnamefont {R.~W.}\ \bibnamefont
  {Schoenlein}}, \ and\ \bibinfo {author} {\bibfnamefont {R.~A.}\ \bibnamefont
  {Kaindl}},\ }\bibfield  {title} {\enquote {\bibinfo {title} {{Ultrafast
  charge localization in a stripe-phase nickelate}},}\ }\href@noop {}
  {\bibfield  {journal} {\bibinfo  {journal} {Nat. Commun.}\ }\textbf {\bibinfo
  {volume} {4}},\ \bibinfo {pages} {2643} (\bibinfo {year} {2013})}\BibitemShut
  {NoStop}%
\bibitem [{\citenamefont {Emery}\ and\ \citenamefont
  {Kivelson}(1995)}]{emer95b}%
  \BibitemOpen
  \bibfield  {author} {\bibinfo {author} {\bibfnamefont {V.~J.}\ \bibnamefont
  {Emery}}\ and\ \bibinfo {author} {\bibfnamefont {S.~A.}\ \bibnamefont
  {Kivelson}},\ }\bibfield  {title} {\enquote {\bibinfo {title}
  {Superconductivity in bad metals},}\ }\href@noop {} {\bibfield  {journal}
  {\bibinfo  {journal} {Phys. Rev. Lett.}\ }\textbf {\bibinfo {volume} {74}},\
  \bibinfo {pages} {3253} (\bibinfo {year} {1995})}\BibitemShut {NoStop}%
\bibitem [{\citenamefont {Ido}\ \emph {et~al.}(1991)\citenamefont {Ido},
  \citenamefont {Magoshi}, \citenamefont {Eisaki},\ and\ \citenamefont
  {Uchida}}]{ido91}%
  \BibitemOpen
  \bibfield  {author} {\bibinfo {author} {\bibfnamefont {T.}~\bibnamefont
  {Ido}}, \bibinfo {author} {\bibfnamefont {K.}~\bibnamefont {Magoshi}},
  \bibinfo {author} {\bibfnamefont {H.}~\bibnamefont {Eisaki}}, \ and\ \bibinfo
  {author} {\bibfnamefont {S.}~\bibnamefont {Uchida}},\ }\bibfield  {title}
  {\enquote {\bibinfo {title} {{Optical study of the
  ${\mathrm{La}}_{2\mathrm{\ensuremath{-}}\mathit{x}}$${\mathrm{Sr}}_{\mathit{x}}$${\mathrm{NiO}}_{4}$
  system: Effect of hole doping on the electronic structure of the
  ${\mathrm{NiO}}_{2}$ plane}},}\ }\href {\doibase 10.1103/PhysRevB.44.12094}
  {\bibfield  {journal} {\bibinfo  {journal} {Phys. Rev. B}\ }\textbf {\bibinfo
  {volume} {44}},\ \bibinfo {pages} {12094--12097} (\bibinfo {year}
  {1991})}\BibitemShut {NoStop}%
\bibitem [{\citenamefont {Berlinsky}(1979)}]{berl79}%
  \BibitemOpen
  \bibfield  {author} {\bibinfo {author} {\bibfnamefont {A.~J.}\ \bibnamefont
  {Berlinsky}},\ }\bibfield  {title} {\enquote {\bibinfo {title}
  {{One-dimensional metals and charge density wave effects in these
  materials}},}\ }\href {http://stacks.iop.org/0034-4885/42/i=7/a=004}
  {\bibfield  {journal} {\bibinfo  {journal} {Rep. Prog. Phys.}\ }\textbf
  {\bibinfo {volume} {42}},\ \bibinfo {pages} {1243} (\bibinfo {year}
  {1979})}\BibitemShut {NoStop}%
\bibitem [{\citenamefont {Gr\"uner}(1988)}]{grun88}%
  \BibitemOpen
  \bibfield  {author} {\bibinfo {author} {\bibfnamefont {G.}~\bibnamefont
  {Gr\"uner}},\ }\bibfield  {title} {\enquote {\bibinfo {title} {{The dynamics
  of charge-density waves}},}\ }\href@noop {} {\bibfield  {journal} {\bibinfo
  {journal} {Rev. Mod. Phys.}\ }\textbf {\bibinfo {volume} {60}},\ \bibinfo
  {pages} {1129--1181} (\bibinfo {year} {1988})}\BibitemShut {NoStop}%
\bibitem [{\citenamefont {Lee}\ \emph {et~al.}(1973)\citenamefont {Lee},
  \citenamefont {Rice},\ and\ \citenamefont {Anderson}}]{lee73}%
  \BibitemOpen
  \bibfield  {author} {\bibinfo {author} {\bibfnamefont {P.~A.}\ \bibnamefont
  {Lee}}, \bibinfo {author} {\bibfnamefont {T.~M.}\ \bibnamefont {Rice}}, \
  and\ \bibinfo {author} {\bibfnamefont {P.~W.}\ \bibnamefont {Anderson}},\
  }\bibfield  {title} {\enquote {\bibinfo {title} {{Fluctuation Effects at a
  Peierls Transition}},}\ }\href {\doibase 10.1103/PhysRevLett.31.462}
  {\bibfield  {journal} {\bibinfo  {journal} {Phys. Rev. Lett.}\ }\textbf
  {\bibinfo {volume} {31}},\ \bibinfo {pages} {462--465} (\bibinfo {year}
  {1973})}\BibitemShut {NoStop}%
\bibitem [{\citenamefont {Anisimov}\ \emph {et~al.}(1992)\citenamefont
  {Anisimov}, \citenamefont {Korotin}, \citenamefont {Zaanen},\ and\
  \citenamefont {Andersen}}]{anis92}%
  \BibitemOpen
  \bibfield  {author} {\bibinfo {author} {\bibfnamefont {V.~I.}\ \bibnamefont
  {Anisimov}}, \bibinfo {author} {\bibfnamefont {M.~A.}\ \bibnamefont
  {Korotin}}, \bibinfo {author} {\bibfnamefont {J.}~\bibnamefont {Zaanen}}, \
  and\ \bibinfo {author} {\bibfnamefont {O.~K.}\ \bibnamefont {Andersen}},\
  }\bibfield  {title} {\enquote {\bibinfo {title} {{Spin bags, polarons, and
  impurity potentials in
  ${\mathrm{La}}_{2\mathrm{-}\mathit{x}}$${\mathrm{Sr}}_{\mathit{x}}$${\mathrm{CuO}}_{4}$
  from first principles}},}\ }\href {\doibase 10.1103/PhysRevLett.68.345}
  {\bibfield  {journal} {\bibinfo  {journal} {Phys. Rev. Lett.}\ }\textbf
  {\bibinfo {volume} {68}},\ \bibinfo {pages} {345--348} (\bibinfo {year}
  {1992})}\BibitemShut {NoStop}%
\bibitem [{\citenamefont {Chen}\ \emph {et~al.}(1993)\citenamefont {Chen},
  \citenamefont {Cheong},\ and\ \citenamefont {Cooper}}]{chen93}%
  \BibitemOpen
  \bibfield  {author} {\bibinfo {author} {\bibfnamefont {C.~H.}\ \bibnamefont
  {Chen}}, \bibinfo {author} {\bibfnamefont {S-W.}\ \bibnamefont {Cheong}}, \
  and\ \bibinfo {author} {\bibfnamefont {A.~S.}\ \bibnamefont {Cooper}},\
  }\bibfield  {title} {\enquote {\bibinfo {title} {{Charge modulations in
  ${\mathrm{La}}_{2\mathrm{-}\mathit{x}}$${\mathrm{Sr}}_{\mathit{x}}$${\mathrm{NiO}}_{4+\mathit{y}}$:
  Ordering of polarons}},}\ }\href {\doibase 10.1103/PhysRevLett.71.2461}
  {\bibfield  {journal} {\bibinfo  {journal} {Phys. Rev. Lett.}\ }\textbf
  {\bibinfo {volume} {71}},\ \bibinfo {pages} {2461--2464} (\bibinfo {year}
  {1993})}\BibitemShut {NoStop}%
\bibitem [{\citenamefont {Fehske}\ \emph {et~al.}(1995)\citenamefont {Fehske},
  \citenamefont {R\"oder}, \citenamefont {Wellein},\ and\ \citenamefont
  {Mistriotis}}]{fehs95}%
  \BibitemOpen
  \bibfield  {author} {\bibinfo {author} {\bibfnamefont {H.}~\bibnamefont
  {Fehske}}, \bibinfo {author} {\bibfnamefont {H.}~\bibnamefont {R\"oder}},
  \bibinfo {author} {\bibfnamefont {G.}~\bibnamefont {Wellein}}, \ and\
  \bibinfo {author} {\bibfnamefont {A.}~\bibnamefont {Mistriotis}},\ }\bibfield
   {title} {\enquote {\bibinfo {title} {{Hole-polaron formation in the
  two-dimensional Holstein \textit{t} - \textit{J} model: A variational Lanczos
  study}},}\ }\href {\doibase 10.1103/PhysRevB.51.16582} {\bibfield  {journal}
  {\bibinfo  {journal} {Phys. Rev. B}\ }\textbf {\bibinfo {volume} {51}},\
  \bibinfo {pages} {16582--16593} (\bibinfo {year} {1995})}\BibitemShut
  {NoStop}%
\bibitem [{\citenamefont {B\"auml}\ \emph {et~al.}(1998)\citenamefont
  {B\"auml}, \citenamefont {Wellein},\ and\ \citenamefont {Fehske}}]{baum98}%
  \BibitemOpen
  \bibfield  {author} {\bibinfo {author} {\bibfnamefont {B.}~\bibnamefont
  {B\"auml}}, \bibinfo {author} {\bibfnamefont {G.}~\bibnamefont {Wellein}}, \
  and\ \bibinfo {author} {\bibfnamefont {H.}~\bibnamefont {Fehske}},\
  }\bibfield  {title} {\enquote {\bibinfo {title} {{Optical absorption and
  single-particle excitations in the two-dimensional Holstein $t-J$ model}},}\
  }\href {\doibase 10.1103/PhysRevB.58.3663} {\bibfield  {journal} {\bibinfo
  {journal} {Phys. Rev. B}\ }\textbf {\bibinfo {volume} {58}},\ \bibinfo
  {pages} {3663--3676} (\bibinfo {year} {1998})}\BibitemShut {NoStop}%
\bibitem [{\citenamefont {Maschek}\ \emph {et~al.}(2016)\citenamefont
  {Maschek}, \citenamefont {Lamago}, \citenamefont {Castellan}, \citenamefont
  {Bosak}, \citenamefont {Reznik},\ and\ \citenamefont {Weber}}]{masc16}%
  \BibitemOpen
  \bibfield  {author} {\bibinfo {author} {\bibfnamefont {M.}~\bibnamefont
  {Maschek}}, \bibinfo {author} {\bibfnamefont {D.}~\bibnamefont {Lamago}},
  \bibinfo {author} {\bibfnamefont {J.-P.}\ \bibnamefont {Castellan}}, \bibinfo
  {author} {\bibfnamefont {A.}~\bibnamefont {Bosak}}, \bibinfo {author}
  {\bibfnamefont {D.}~\bibnamefont {Reznik}}, \ and\ \bibinfo {author}
  {\bibfnamefont {F.}~\bibnamefont {Weber}},\ }\bibfield  {title} {\enquote
  {\bibinfo {title} {{Polaronic metal phases in
  $\mathrm{L}{\mathrm{a}}_{0.7}\mathrm{S}{\mathrm{r}}_{0.3}\mathrm{Mn}{\mathrm{O}}_{3}$
  uncovered by inelastic neutron and x-ray scattering}},}\ }\href {\doibase
  10.1103/PhysRevB.93.045112} {\bibfield  {journal} {\bibinfo  {journal} {Phys.
  Rev. B}\ }\textbf {\bibinfo {volume} {93}},\ \bibinfo {pages} {045112}
  (\bibinfo {year} {2016})}\BibitemShut {NoStop}%
\bibitem [{\citenamefont {Abeykoon}\ \emph {et~al.}(2013)\citenamefont
  {Abeykoon}, \citenamefont {Bo\ifmmode~\check{z}\else \v{z}\fi{}in},
  \citenamefont {Yin}, \citenamefont {Gu}, \citenamefont {Hill}, \citenamefont
  {Tranquada},\ and\ \citenamefont {Billinge}}]{abey13}%
  \BibitemOpen
  \bibfield  {author} {\bibinfo {author} {\bibfnamefont {A.~M.~Milinda}\
  \bibnamefont {Abeykoon}}, \bibinfo {author} {\bibfnamefont {Emil~S.}\
  \bibnamefont {Bo\ifmmode~\check{z}\else \v{z}\fi{}in}}, \bibinfo {author}
  {\bibfnamefont {Wei-Guo}\ \bibnamefont {Yin}}, \bibinfo {author}
  {\bibfnamefont {Genda}\ \bibnamefont {Gu}}, \bibinfo {author} {\bibfnamefont
  {John~P.}\ \bibnamefont {Hill}}, \bibinfo {author} {\bibfnamefont {John~M.}\
  \bibnamefont {Tranquada}}, \ and\ \bibinfo {author} {\bibfnamefont {Simon
  J.~L.}\ \bibnamefont {Billinge}},\ }\bibfield  {title} {\enquote {\bibinfo
  {title} {{Evidence for Short-Range-Ordered Charge Stripes Far above the
  Charge-Ordering Transition in La$_{1.67}$Sr$_{0.33}$NiO$_{4}$}},}\
  }\href@noop {} {\bibfield  {journal} {\bibinfo  {journal} {Phys. Rev. Lett.}\
  }\textbf {\bibinfo {volume} {111}},\ \bibinfo {pages} {096404} (\bibinfo
  {year} {2013})}\BibitemShut {NoStop}%
\bibitem [{\citenamefont {Anissimova}\ \emph {et~al.}(2014)\citenamefont
  {Anissimova}, \citenamefont {Parshall}, \citenamefont {Gu}, \citenamefont
  {Marty}, \citenamefont {Lumsden}, \citenamefont {Chi}, \citenamefont
  {Fernandez-Baca}, \citenamefont {Abernathy}, \citenamefont {Lamago},
  \citenamefont {Tranquada},\ and\ \citenamefont {Reznik}}]{anis14}%
  \BibitemOpen
  \bibfield  {author} {\bibinfo {author} {\bibfnamefont {S.}~\bibnamefont
  {Anissimova}}, \bibinfo {author} {\bibfnamefont {D.}~\bibnamefont
  {Parshall}}, \bibinfo {author} {\bibfnamefont {G.~D.}\ \bibnamefont {Gu}},
  \bibinfo {author} {\bibfnamefont {K.}~\bibnamefont {Marty}}, \bibinfo
  {author} {\bibfnamefont {M.~D.}\ \bibnamefont {Lumsden}}, \bibinfo {author}
  {\bibfnamefont {Songxue}\ \bibnamefont {Chi}}, \bibinfo {author}
  {\bibfnamefont {J.~A.}\ \bibnamefont {Fernandez-Baca}}, \bibinfo {author}
  {\bibfnamefont {D.~L.}\ \bibnamefont {Abernathy}}, \bibinfo {author}
  {\bibfnamefont {D.}~\bibnamefont {Lamago}}, \bibinfo {author} {\bibfnamefont
  {J.~M.}\ \bibnamefont {Tranquada}}, \ and\ \bibinfo {author} {\bibfnamefont
  {D.}~\bibnamefont {Reznik}},\ }\bibfield  {title} {\enquote {\bibinfo {title}
  {{Direct observation of dynamic charge stripes in
  La$_{2-x}$Sr$_x$NiO$_4$}},}\ }\href@noop {} {\bibfield  {journal} {\bibinfo
  {journal} {Nat. Commun.}\ }\textbf {\bibinfo {volume} {5}},\ \bibinfo {pages}
  {3467} (\bibinfo {year} {2014})}\BibitemShut {NoStop}%
\bibitem [{\citenamefont {Kivelson}\ \emph {et~al.}(1998)\citenamefont
  {Kivelson}, \citenamefont {Fradkin},\ and\ \citenamefont {Emery}}]{kive98}%
  \BibitemOpen
  \bibfield  {author} {\bibinfo {author} {\bibfnamefont {S.~A.}\ \bibnamefont
  {Kivelson}}, \bibinfo {author} {\bibfnamefont {E.}~\bibnamefont {Fradkin}}, \
  and\ \bibinfo {author} {\bibfnamefont {V.~J.}\ \bibnamefont {Emery}},\
  }\bibfield  {title} {\enquote {\bibinfo {title} {{Electronic liquid-crystal
  phases of a doped Mott insulator}},}\ }\href@noop {} {\bibfield  {journal}
  {\bibinfo  {journal} {Nature}\ }\textbf {\bibinfo {volume} {393}},\ \bibinfo
  {pages} {550--553} (\bibinfo {year} {1998})}\BibitemShut {NoStop}%
\bibitem [{\citenamefont {Fradkin}\ \emph {et~al.}(2010)\citenamefont
  {Fradkin}, \citenamefont {Kivelson}, \citenamefont {Lawler}, \citenamefont
  {Eisenstein},\ and\ \citenamefont {Mackenzie}}]{frad10}%
  \BibitemOpen
  \bibfield  {author} {\bibinfo {author} {\bibfnamefont {Eduardo}\ \bibnamefont
  {Fradkin}}, \bibinfo {author} {\bibfnamefont {Steven~A.}\ \bibnamefont
  {Kivelson}}, \bibinfo {author} {\bibfnamefont {Michael~J.}\ \bibnamefont
  {Lawler}}, \bibinfo {author} {\bibfnamefont {James~P.}\ \bibnamefont
  {Eisenstein}}, \ and\ \bibinfo {author} {\bibfnamefont {Andrew~P.}\
  \bibnamefont {Mackenzie}},\ }\bibfield  {title} {\enquote {\bibinfo {title}
  {{Nematic Fermi Fluids in Condensed Matter Physics}},}\ }\href@noop {}
  {\bibfield  {journal} {\bibinfo  {journal} {Annu. Rev. Condens. Matter
  Phys.}\ }\textbf {\bibinfo {volume} {1}},\ \bibinfo {pages} {153--178}
  (\bibinfo {year} {2010})}\BibitemShut {NoStop}%
\bibitem [{\citenamefont {Nie}\ \emph {et~al.}(2014)\citenamefont {Nie},
  \citenamefont {Tarjus},\ and\ \citenamefont {Kivelson}}]{nie14}%
  \BibitemOpen
  \bibfield  {author} {\bibinfo {author} {\bibfnamefont {Laimei}\ \bibnamefont
  {Nie}}, \bibinfo {author} {\bibfnamefont {Gilles}\ \bibnamefont {Tarjus}}, \
  and\ \bibinfo {author} {\bibfnamefont {Steven~Allan}\ \bibnamefont
  {Kivelson}},\ }\bibfield  {title} {\enquote {\bibinfo {title} {{Quenched
  disorder and vestigial nematicity in the pseudogap regime of the
  cuprates}},}\ }\href@noop {} {\bibfield  {journal} {\bibinfo  {journal}
  {Proc. Natl. Acad. Sci. USA}\ }\textbf {\bibinfo {volume} {111}},\ \bibinfo
  {pages} {7980--7985} (\bibinfo {year} {2014})}\BibitemShut {NoStop}%
\bibitem [{\citenamefont {H\"ucker}\ \emph {et~al.}(2004)\citenamefont
  {H\"ucker}, \citenamefont {Chung}, \citenamefont {Chand}, \citenamefont
  {Vogt}, \citenamefont {Tranquada},\ and\ \citenamefont {Buttrey}}]{huck04b}%
  \BibitemOpen
  \bibfield  {author} {\bibinfo {author} {\bibfnamefont {M.}~\bibnamefont
  {H\"ucker}}, \bibinfo {author} {\bibfnamefont {K.}~\bibnamefont {Chung}},
  \bibinfo {author} {\bibfnamefont {M.}~\bibnamefont {Chand}}, \bibinfo
  {author} {\bibfnamefont {T.}~\bibnamefont {Vogt}}, \bibinfo {author}
  {\bibfnamefont {J.~M.}\ \bibnamefont {Tranquada}}, \ and\ \bibinfo {author}
  {\bibfnamefont {D.~J.}\ \bibnamefont {Buttrey}},\ }\bibfield  {title}
  {\enquote {\bibinfo {title} {{Oxygen and strontium codoping of La$_2$NiO$_4$:
  Room-temperature phase diagrams}},}\ }\href@noop {} {\bibfield  {journal}
  {\bibinfo  {journal} {Phys. Rev. B}\ }\textbf {\bibinfo {volume} {70}},\
  \bibinfo {pages} {064105} (\bibinfo {year} {2004})}\BibitemShut {NoStop}%
\bibitem [{\citenamefont {Winn}\ \emph {et~al.}(2015)\citenamefont {Winn},
  \citenamefont {Filges}, \citenamefont {Garlea}, \citenamefont {Graves-Brook},
  \citenamefont {Hagen}, \citenamefont {Jiang}, \citenamefont {Kenzelmann},
  \citenamefont {Passell}, \citenamefont {Shapiro}, \citenamefont {Tong},\ and\
  \citenamefont {Zaliznyak}}]{hyspec15}%
  \BibitemOpen
  \bibfield  {author} {\bibinfo {author} {\bibfnamefont {B.}~\bibnamefont
  {Winn}}, \bibinfo {author} {\bibfnamefont {U.}~\bibnamefont {Filges}},
  \bibinfo {author} {\bibfnamefont {O.~V.}\ \bibnamefont {Garlea}}, \bibinfo
  {author} {\bibfnamefont {M.}~\bibnamefont {Graves-Brook}}, \bibinfo {author}
  {\bibfnamefont {M.}~\bibnamefont {Hagen}}, \bibinfo {author} {\bibfnamefont
  {C.~Y.}\ \bibnamefont {Jiang}}, \bibinfo {author} {\bibfnamefont
  {M.}~\bibnamefont {Kenzelmann}}, \bibinfo {author} {\bibfnamefont
  {L.}~\bibnamefont {Passell}}, \bibinfo {author} {\bibfnamefont {S.~M.}\
  \bibnamefont {Shapiro}}, \bibinfo {author} {\bibfnamefont {X.}~\bibnamefont
  {Tong}}, \ and\ \bibinfo {author} {\bibfnamefont {I.}~\bibnamefont
  {Zaliznyak}},\ }\bibfield  {title} {\enquote {\bibinfo {title} {{Recent
  progress on HYSPEC, and its polarization analysis capabilities}},}\ }\href
  {http://dx.doi.org/10.1051/epjconf/20158303017} {\bibfield  {journal}
  {\bibinfo  {journal} {EPJ Web Conf.}\ }\textbf {\bibinfo {volume} {83}},\
  \bibinfo {pages} {03017} (\bibinfo {year} {2015})}\BibitemShut {NoStop}%
\bibitem [{\citenamefont {Arnold}\ \emph {et~al.}(2014)\citenamefont {Arnold},
  \citenamefont {Bilheux}, \citenamefont {Borreguero}, \citenamefont {Buts},
  \citenamefont {Campbell}, \citenamefont {Chapon}, \citenamefont {Doucet},
  \citenamefont {Draper}, \citenamefont {Leal}, \citenamefont {Gigg},
  \citenamefont {Lynch}, \citenamefont {Markvardsen}, \citenamefont
  {Mikkelson}, \citenamefont {Mikkelson}, \citenamefont {Miller}, \citenamefont
  {Palmen}, \citenamefont {Parker}, \citenamefont {Passos}, \citenamefont
  {Perring}, \citenamefont {Peterson}, \citenamefont {Ren}, \citenamefont
  {Reuter}, \citenamefont {Savici}, \citenamefont {Taylor}, \citenamefont
  {Taylor}, \citenamefont {Tolchenov}, \citenamefont {Zhou},\ and\
  \citenamefont {Zikovsky}}]{mantid14}%
  \BibitemOpen
  \bibfield  {author} {\bibinfo {author} {\bibfnamefont {O.}~\bibnamefont
  {Arnold}}, \bibinfo {author} {\bibfnamefont {J.C.}\ \bibnamefont {Bilheux}},
  \bibinfo {author} {\bibfnamefont {J.M.}\ \bibnamefont {Borreguero}}, \bibinfo
  {author} {\bibfnamefont {A.}~\bibnamefont {Buts}}, \bibinfo {author}
  {\bibfnamefont {S.I.}\ \bibnamefont {Campbell}}, \bibinfo {author}
  {\bibfnamefont {L.}~\bibnamefont {Chapon}}, \bibinfo {author} {\bibfnamefont
  {M.}~\bibnamefont {Doucet}}, \bibinfo {author} {\bibfnamefont
  {N.}~\bibnamefont {Draper}}, \bibinfo {author} {\bibfnamefont {R.~Ferraz}\
  \bibnamefont {Leal}}, \bibinfo {author} {\bibfnamefont {M.A.}\ \bibnamefont
  {Gigg}}, \bibinfo {author} {\bibfnamefont {V.E.}\ \bibnamefont {Lynch}},
  \bibinfo {author} {\bibfnamefont {A.}~\bibnamefont {Markvardsen}}, \bibinfo
  {author} {\bibfnamefont {D.J.}\ \bibnamefont {Mikkelson}}, \bibinfo {author}
  {\bibfnamefont {R.L.}\ \bibnamefont {Mikkelson}}, \bibinfo {author}
  {\bibfnamefont {R.}~\bibnamefont {Miller}}, \bibinfo {author} {\bibfnamefont
  {K.}~\bibnamefont {Palmen}}, \bibinfo {author} {\bibfnamefont
  {P.}~\bibnamefont {Parker}}, \bibinfo {author} {\bibfnamefont
  {G.}~\bibnamefont {Passos}}, \bibinfo {author} {\bibfnamefont {T.G.}\
  \bibnamefont {Perring}}, \bibinfo {author} {\bibfnamefont {P.F.}\
  \bibnamefont {Peterson}}, \bibinfo {author} {\bibfnamefont {S.}~\bibnamefont
  {Ren}}, \bibinfo {author} {\bibfnamefont {M.A.}\ \bibnamefont {Reuter}},
  \bibinfo {author} {\bibfnamefont {A.T.}\ \bibnamefont {Savici}}, \bibinfo
  {author} {\bibfnamefont {J.W.}\ \bibnamefont {Taylor}}, \bibinfo {author}
  {\bibfnamefont {R.J.}\ \bibnamefont {Taylor}}, \bibinfo {author}
  {\bibfnamefont {R.}~\bibnamefont {Tolchenov}}, \bibinfo {author}
  {\bibfnamefont {W.}~\bibnamefont {Zhou}}, \ and\ \bibinfo {author}
  {\bibfnamefont {J.}~\bibnamefont {Zikovsky}},\ }\bibfield  {title} {\enquote
  {\bibinfo {title} {{Mantid---Data analysis and visualization package for
  neutron scattering and $\mu$SR experiments}},}\ }\href@noop {} {\bibfield
  {journal} {\bibinfo  {journal} {Nucl. Instrum. Methods A}\ }\textbf {\bibinfo
  {volume} {764}},\ \bibinfo {pages} {156--166} (\bibinfo {year}
  {2014})}\BibitemShut {NoStop}%
\bibitem [{\citenamefont {Azuah}\ \emph {et~al.}(2009)\citenamefont {Azuah},
  \citenamefont {Kneller}, \citenamefont {Qiu}, \citenamefont
  {Tregenna-Piggott}, \citenamefont {Brown}, \citenamefont {Copley},\ and\
  \citenamefont {Dimeo}}]{dave09}%
  \BibitemOpen
  \bibfield  {author} {\bibinfo {author} {\bibfnamefont {R.~T.}\ \bibnamefont
  {Azuah}}, \bibinfo {author} {\bibfnamefont {L.~R.}\ \bibnamefont {Kneller}},
  \bibinfo {author} {\bibfnamefont {Y.~M.}\ \bibnamefont {Qiu}}, \bibinfo
  {author} {\bibfnamefont {P.~L.~W.}\ \bibnamefont {Tregenna-Piggott}},
  \bibinfo {author} {\bibfnamefont {C.~M.}\ \bibnamefont {Brown}}, \bibinfo
  {author} {\bibfnamefont {J.~R.~D.}\ \bibnamefont {Copley}}, \ and\ \bibinfo
  {author} {\bibfnamefont {R.~M.}\ \bibnamefont {Dimeo}},\ }\bibfield  {title}
  {\enquote {\bibinfo {title} {{DAVE: A Comprehensive Software Suite for the
  Reduction, Visualization, and Analysis of Low Energy Neutron Spectroscopic
  Data}},}\ }\href
  {http://nvlpubs.nist.gov/nistpubs/jres/114/6/V114.N06.A04.pdf} {\bibfield
  {journal} {\bibinfo  {journal} {J. Res. Natl. Inst. Stan. Technol.}\ }\textbf
  {\bibinfo {volume} {114}},\ \bibinfo {pages} {341} (\bibinfo {year}
  {2009})}\BibitemShut {NoStop}%
\bibitem [{sup()}]{suppl}%
  \BibitemOpen
  \href@noop {} {}\bibinfo {howpublished} {{Supplemental Material is available
  at the end of this file.}}\BibitemShut {Stop}%
\bibitem [{\citenamefont {Li}\ \emph {et~al.}(2003)\citenamefont {Li},
  \citenamefont {Zhu}, \citenamefont {Tranquada}, \citenamefont {Yamada},\ and\
  \citenamefont {Buttrey}}]{li03}%
  \BibitemOpen
  \bibfield  {author} {\bibinfo {author} {\bibfnamefont {Jianqi}\ \bibnamefont
  {Li}}, \bibinfo {author} {\bibfnamefont {Yimei}\ \bibnamefont {Zhu}},
  \bibinfo {author} {\bibfnamefont {J.~M.}\ \bibnamefont {Tranquada}}, \bibinfo
  {author} {\bibfnamefont {K.}~\bibnamefont {Yamada}}, \ and\ \bibinfo {author}
  {\bibfnamefont {D.~J.}\ \bibnamefont {Buttrey}},\ }\bibfield  {title}
  {\enquote {\bibinfo {title} {{Transmission-electron-microscopy study of
  charge-stripe order in
  ${\mathrm{La}}_{1.725}{\mathrm{Sr}}_{0.275}{\mathrm{NiO}}_{4}$}},}\ }\href
  {\doibase 10.1103/PhysRevB.67.012404} {\bibfield  {journal} {\bibinfo
  {journal} {Phys. Rev. B}\ }\textbf {\bibinfo {volume} {67}},\ \bibinfo
  {pages} {012404} (\bibinfo {year} {2003})}\BibitemShut {NoStop}%
\bibitem [{\citenamefont {Hucker}\ \emph {et~al.}(2006)\citenamefont {Hucker},
  \citenamefont {v.~Zimmermann}, \citenamefont {Klingeler}, \citenamefont
  {Kiele}, \citenamefont {Geck}, \citenamefont {Bakehe}, \citenamefont {Zhang},
  \citenamefont {Hill}, \citenamefont {Revcolevschi}, \citenamefont {Buttrey},
  \citenamefont {Buchner},\ and\ \citenamefont {Tranquada}}]{huck06}%
  \BibitemOpen
  \bibfield  {author} {\bibinfo {author} {\bibfnamefont {M.}~\bibnamefont
  {Hucker}}, \bibinfo {author} {\bibfnamefont {M.}~\bibnamefont
  {v.~Zimmermann}}, \bibinfo {author} {\bibfnamefont {R.}~\bibnamefont
  {Klingeler}}, \bibinfo {author} {\bibfnamefont {S.}~\bibnamefont {Kiele}},
  \bibinfo {author} {\bibfnamefont {J.}~\bibnamefont {Geck}}, \bibinfo {author}
  {\bibfnamefont {S.~N.}\ \bibnamefont {Bakehe}}, \bibinfo {author}
  {\bibfnamefont {J.~Z.}\ \bibnamefont {Zhang}}, \bibinfo {author}
  {\bibfnamefont {J.~P.}\ \bibnamefont {Hill}}, \bibinfo {author}
  {\bibfnamefont {A.}~\bibnamefont {Revcolevschi}}, \bibinfo {author}
  {\bibfnamefont {D.~J.}\ \bibnamefont {Buttrey}}, \bibinfo {author}
  {\bibfnamefont {B.}~\bibnamefont {Buchner}}, \ and\ \bibinfo {author}
  {\bibfnamefont {J.~M.}\ \bibnamefont {Tranquada}},\ }\bibfield  {title}
  {\enquote {\bibinfo {title} {{Unidirectional diagonal order and
  three-dimensional stacking of charge stripes in orthorhombic
  Pr$_{1.67}$Sr$_{0.33}$NiO$_4$ and Nd$_{1.67}$Sr$_{0.33}$NiO$_4$}},}\
  }\href@noop {} {\bibfield  {journal} {\bibinfo  {journal} {Phys. Rev. B}\
  }\textbf {\bibinfo {volume} {74}},\ \bibinfo {eid} {085112} (\bibinfo {year}
  {2006})}\BibitemShut {NoStop}%
\bibitem [{\citenamefont {Tranquada}\ \emph {et~al.}(1996)\citenamefont
  {Tranquada}, \citenamefont {Buttrey},\ and\ \citenamefont
  {Sachan}}]{tran96a}%
  \BibitemOpen
  \bibfield  {author} {\bibinfo {author} {\bibfnamefont {J.~M.}\ \bibnamefont
  {Tranquada}}, \bibinfo {author} {\bibfnamefont {D.~J.}\ \bibnamefont
  {Buttrey}}, \ and\ \bibinfo {author} {\bibfnamefont {V.}~\bibnamefont
  {Sachan}},\ }\bibfield  {title} {\enquote {\bibinfo {title} {{Incommensurate
  stripe order in La$_{2-x}$Sr$_{x}$NiO$_{4}$ with $x =0.225$}},}\ }\href@noop
  {} {\bibfield  {journal} {\bibinfo  {journal} {Phys. Rev. B}\ }\textbf
  {\bibinfo {volume} {54}},\ \bibinfo {pages} {12318--12323} (\bibinfo {year}
  {1996})}\BibitemShut {NoStop}%
\bibitem [{\citenamefont {Lee}\ \emph {et~al.}(2002)\citenamefont {Lee},
  \citenamefont {Tranquada}, \citenamefont {Yamada}, \citenamefont {Buttrey},
  \citenamefont {Li},\ and\ \citenamefont {Cheong}}]{lee02}%
  \BibitemOpen
  \bibfield  {author} {\bibinfo {author} {\bibfnamefont {S.-H.}\ \bibnamefont
  {Lee}}, \bibinfo {author} {\bibfnamefont {J.~M.}\ \bibnamefont {Tranquada}},
  \bibinfo {author} {\bibfnamefont {K.}~\bibnamefont {Yamada}}, \bibinfo
  {author} {\bibfnamefont {D.~J.}\ \bibnamefont {Buttrey}}, \bibinfo {author}
  {\bibfnamefont {Q.}~\bibnamefont {Li}}, \ and\ \bibinfo {author}
  {\bibfnamefont {S.-W.}\ \bibnamefont {Cheong}},\ }\bibfield  {title}
  {\enquote {\bibinfo {title} {{Freezing of a Stripe Liquid}},}\ }\href@noop {}
  {\bibfield  {journal} {\bibinfo  {journal} {Phys. Rev. Lett.}\ }\textbf
  {\bibinfo {volume} {88}},\ \bibinfo {pages} {126401} (\bibinfo {year}
  {2002})}\BibitemShut {NoStop}%
\bibitem [{\citenamefont {Spencer}\ \emph {et~al.}(2005)\citenamefont
  {Spencer}, \citenamefont {Ghazi}, \citenamefont {Wilkins}, \citenamefont
  {Hatton}, \citenamefont {Brown}, \citenamefont {Prabhakaran},\ and\
  \citenamefont {Boothroyd}}]{spen05}%
  \BibitemOpen
  \bibfield  {author} {\bibinfo {author} {\bibfnamefont {P.~D.}\ \bibnamefont
  {Spencer}}, \bibinfo {author} {\bibfnamefont {M.~E.}\ \bibnamefont {Ghazi}},
  \bibinfo {author} {\bibfnamefont {S.~B.}\ \bibnamefont {Wilkins}}, \bibinfo
  {author} {\bibfnamefont {P.~D.}\ \bibnamefont {Hatton}}, \bibinfo {author}
  {\bibfnamefont {S.~D.}\ \bibnamefont {Brown}}, \bibinfo {author}
  {\bibfnamefont {D.}~\bibnamefont {Prabhakaran}}, \ and\ \bibinfo {author}
  {\bibfnamefont {A.~T.}\ \bibnamefont {Boothroyd}},\ }\bibfield  {title}
  {\enquote {\bibinfo {title} {{Charge stripe glasses in
  La$_{2-x}$Sr$_x$NiO$_4$ for $0.20 < x < 0.25$}},}\ }\href@noop {} {\bibfield
  {journal} {\bibinfo  {journal} {Eur. Phys. J. B}\ }\textbf {\bibinfo {volume}
  {46}},\ \bibinfo {pages} {27--32} (\bibinfo {year} {2005})}\BibitemShut
  {NoStop}%
\bibitem [{\citenamefont {Lee}\ \emph {et~al.}(2012)\citenamefont {Lee},
  \citenamefont {Chuang}, \citenamefont {Moore}, \citenamefont {Zhu},
  \citenamefont {Patthey}, \citenamefont {Trigo}, \citenamefont {Lu},
  \citenamefont {Kirchmann}, \citenamefont {Krupin}, \citenamefont {Yi},
  \citenamefont {Langner}, \citenamefont {Huse}, \citenamefont {Robinson},
  \citenamefont {Chen}, \citenamefont {Zhou}, \citenamefont {Coslovich},
  \citenamefont {Huber}, \citenamefont {Reis}, \citenamefont {Kaindl},
  \citenamefont {Schoenlein}, \citenamefont {Doering}, \citenamefont {Denes},
  \citenamefont {Schlotter}, \citenamefont {Turner}, \citenamefont {Johnson},
  \citenamefont {F{\~A}¶rst}, \citenamefont {Sasagawa}, \citenamefont {Kung},
  \citenamefont {Sorini}, \citenamefont {Kemper}, \citenamefont {Moritz},
  \citenamefont {Devereaux}, \citenamefont {Lee}, \citenamefont {Shen},\ and\
  \citenamefont {Hussain}}]{lee12}%
  \BibitemOpen
  \bibfield  {author} {\bibinfo {author} {\bibfnamefont {W.~S.}\ \bibnamefont
  {Lee}}, \bibinfo {author} {\bibfnamefont {Y.~D.}\ \bibnamefont {Chuang}},
  \bibinfo {author} {\bibfnamefont {R.~G.}\ \bibnamefont {Moore}}, \bibinfo
  {author} {\bibfnamefont {Y.}~\bibnamefont {Zhu}}, \bibinfo {author}
  {\bibfnamefont {L.}~\bibnamefont {Patthey}}, \bibinfo {author} {\bibfnamefont
  {M.}~\bibnamefont {Trigo}}, \bibinfo {author} {\bibfnamefont {D.~H.}\
  \bibnamefont {Lu}}, \bibinfo {author} {\bibfnamefont {P.~S.}\ \bibnamefont
  {Kirchmann}}, \bibinfo {author} {\bibfnamefont {O.}~\bibnamefont {Krupin}},
  \bibinfo {author} {\bibfnamefont {M.}~\bibnamefont {Yi}}, \bibinfo {author}
  {\bibfnamefont {M.}~\bibnamefont {Langner}}, \bibinfo {author} {\bibfnamefont
  {N.}~\bibnamefont {Huse}}, \bibinfo {author} {\bibfnamefont {J.~S.}\
  \bibnamefont {Robinson}}, \bibinfo {author} {\bibfnamefont {Y.}~\bibnamefont
  {Chen}}, \bibinfo {author} {\bibfnamefont {S.~Y.}\ \bibnamefont {Zhou}},
  \bibinfo {author} {\bibfnamefont {G.}~\bibnamefont {Coslovich}}, \bibinfo
  {author} {\bibfnamefont {B.}~\bibnamefont {Huber}}, \bibinfo {author}
  {\bibfnamefont {D.~A.}\ \bibnamefont {Reis}}, \bibinfo {author}
  {\bibfnamefont {R.~A.}\ \bibnamefont {Kaindl}}, \bibinfo {author}
  {\bibfnamefont {R.~W.}\ \bibnamefont {Schoenlein}}, \bibinfo {author}
  {\bibfnamefont {D.}~\bibnamefont {Doering}}, \bibinfo {author} {\bibfnamefont
  {P.}~\bibnamefont {Denes}}, \bibinfo {author} {\bibfnamefont {W.~F.}\
  \bibnamefont {Schlotter}}, \bibinfo {author} {\bibfnamefont {J.~J.}\
  \bibnamefont {Turner}}, \bibinfo {author} {\bibfnamefont {S.~L.}\
  \bibnamefont {Johnson}}, \bibinfo {author} {\bibfnamefont {M.}~\bibnamefont
  {F{\~A}¶rst}}, \bibinfo {author} {\bibfnamefont {T.}~\bibnamefont
  {Sasagawa}}, \bibinfo {author} {\bibfnamefont {Y.~F.}\ \bibnamefont {Kung}},
  \bibinfo {author} {\bibfnamefont {A.~P.}\ \bibnamefont {Sorini}}, \bibinfo
  {author} {\bibfnamefont {A.~F.}\ \bibnamefont {Kemper}}, \bibinfo {author}
  {\bibfnamefont {B.}~\bibnamefont {Moritz}}, \bibinfo {author} {\bibfnamefont
  {T.~P.}\ \bibnamefont {Devereaux}}, \bibinfo {author} {\bibfnamefont {D.~H.}\
  \bibnamefont {Lee}}, \bibinfo {author} {\bibfnamefont {Z.~X.}\ \bibnamefont
  {Shen}}, \ and\ \bibinfo {author} {\bibfnamefont {Z.}~\bibnamefont
  {Hussain}},\ }\bibfield  {title} {\enquote {\bibinfo {title} {{Phase
  fluctuations and the absence of topological defects in a photo-excited
  charge-ordered nickelate}},}\ }\href@noop {} {\bibfield  {journal} {\bibinfo
  {journal} {Nat. Commun.}\ }\textbf {\bibinfo {volume} {3}},\ \bibinfo {pages}
  {838} (\bibinfo {year} {2012})}\BibitemShut {NoStop}%
\bibitem [{\citenamefont {Lloyd-Hughes}\ \emph {et~al.}(2008)\citenamefont
  {Lloyd-Hughes}, \citenamefont {Prabhakaran}, \citenamefont {Boothroyd},\ and\
  \citenamefont {Johnston}}]{lloy08}%
  \BibitemOpen
  \bibfield  {author} {\bibinfo {author} {\bibfnamefont {J.}~\bibnamefont
  {Lloyd-Hughes}}, \bibinfo {author} {\bibfnamefont {D.}~\bibnamefont
  {Prabhakaran}}, \bibinfo {author} {\bibfnamefont {A.~T.}\ \bibnamefont
  {Boothroyd}}, \ and\ \bibinfo {author} {\bibfnamefont {M.~B.}\ \bibnamefont
  {Johnston}},\ }\bibfield  {title} {\enquote {\bibinfo {title} {{Low-energy
  collective dynamics of charge stripes in the doped nickelate
  La$_{2-x}$Sr$_{x}$NiO$_{4+\delta}$ observed with optical conductivity
  measurements}},}\ }\href@noop {} {\bibfield  {journal} {\bibinfo  {journal}
  {Phys. Rev. B}\ }\textbf {\bibinfo {volume} {77}},\ \bibinfo {pages} {195114}
  (\bibinfo {year} {2008})}\BibitemShut {NoStop}%
\bibitem [{\citenamefont {Zachar}\ and\ \citenamefont
  {Zaliznyak}(2003)}]{zach03}%
  \BibitemOpen
  \bibfield  {author} {\bibinfo {author} {\bibfnamefont {Oron}\ \bibnamefont
  {Zachar}}\ and\ \bibinfo {author} {\bibfnamefont {Igor}\ \bibnamefont
  {Zaliznyak}},\ }\bibfield  {title} {\enquote {\bibinfo {title} {{Dimensional
  Crossover and Charge Order in Half-Doped Manganites and Cobaltites}},}\
  }\href {\doibase 10.1103/PhysRevLett.91.036401} {\bibfield  {journal}
  {\bibinfo  {journal} {Phys. Rev. Lett.}\ }\textbf {\bibinfo {volume} {91}},\
  \bibinfo {pages} {036401} (\bibinfo {year} {2003})}\BibitemShut {NoStop}%
\bibitem [{\citenamefont {Chaikin}\ and\ \citenamefont
  {Lubensky}(1995)}]{chai95}%
  \BibitemOpen
  \bibfield  {author} {\bibinfo {author} {\bibfnamefont {P.~M.}\ \bibnamefont
  {Chaikin}}\ and\ \bibinfo {author} {\bibfnamefont {T.~C.}\ \bibnamefont
  {Lubensky}},\ }\href@noop {} {\emph {\bibinfo {title} {Principles of
  condensed matter physics}}}\ (\bibinfo  {publisher} {Cambridge University
  Press},\ \bibinfo {address} {Cambridge, UK},\ \bibinfo {year}
  {1995})\BibitemShut {NoStop}%
\bibitem [{\citenamefont {{Le Tacon}}\ \emph {et~al.}(2014)\citenamefont {{Le
  Tacon}}, \citenamefont {Bosak}, \citenamefont {Souliou}, \citenamefont
  {Dellea}, \citenamefont {Loew}, \citenamefont {Heid}, \citenamefont {Bohnen},
  \citenamefont {Ghiringhelli}, \citenamefont {Krisch},\ and\ \citenamefont
  {Keimer}}]{leta14}%
  \BibitemOpen
  \bibfield  {author} {\bibinfo {author} {\bibfnamefont {M.}~\bibnamefont {{Le
  Tacon}}}, \bibinfo {author} {\bibfnamefont {A.}~\bibnamefont {Bosak}},
  \bibinfo {author} {\bibfnamefont {S.~M.}\ \bibnamefont {Souliou}}, \bibinfo
  {author} {\bibfnamefont {G.}~\bibnamefont {Dellea}}, \bibinfo {author}
  {\bibfnamefont {T.}~\bibnamefont {Loew}}, \bibinfo {author} {\bibfnamefont
  {R.}~\bibnamefont {Heid}}, \bibinfo {author} {\bibfnamefont {K-P.}\
  \bibnamefont {Bohnen}}, \bibinfo {author} {\bibfnamefont {G.}~\bibnamefont
  {Ghiringhelli}}, \bibinfo {author} {\bibfnamefont {M.}~\bibnamefont
  {Krisch}}, \ and\ \bibinfo {author} {\bibfnamefont {B.}~\bibnamefont
  {Keimer}},\ }\bibfield  {title} {\enquote {\bibinfo {title} {{Inelastic X-ray
  scattering in YBa$_2$Cu$_3$O$_{6.6}$ reveals giant phonon anomalies and
  elastic central peak due to charge-density-wave formation}},}\ }\href@noop {}
  {\bibfield  {journal} {\bibinfo  {journal} {Nat. Phys.}\ }\textbf {\bibinfo
  {volume} {10}},\ \bibinfo {pages} {52--58} (\bibinfo {year}
  {2014})}\BibitemShut {NoStop}%
\bibitem [{\citenamefont {Blackburn}\ \emph {et~al.}(2013)\citenamefont
  {Blackburn}, \citenamefont {Chang}, \citenamefont {Said}, \citenamefont
  {Leu}, \citenamefont {Liang}, \citenamefont {Bonn}, \citenamefont {Hardy},
  \citenamefont {Forgan},\ and\ \citenamefont {Hayden}}]{blac13b}%
  \BibitemOpen
  \bibfield  {author} {\bibinfo {author} {\bibfnamefont {E.}~\bibnamefont
  {Blackburn}}, \bibinfo {author} {\bibfnamefont {J.}~\bibnamefont {Chang}},
  \bibinfo {author} {\bibfnamefont {A.~H.}\ \bibnamefont {Said}}, \bibinfo
  {author} {\bibfnamefont {B.~M.}\ \bibnamefont {Leu}}, \bibinfo {author}
  {\bibfnamefont {Ruixing}\ \bibnamefont {Liang}}, \bibinfo {author}
  {\bibfnamefont {D.~A.}\ \bibnamefont {Bonn}}, \bibinfo {author}
  {\bibfnamefont {W.~N.}\ \bibnamefont {Hardy}}, \bibinfo {author}
  {\bibfnamefont {E.~M.}\ \bibnamefont {Forgan}}, \ and\ \bibinfo {author}
  {\bibfnamefont {S.~M.}\ \bibnamefont {Hayden}},\ }\bibfield  {title}
  {\enquote {\bibinfo {title} {{Inelastic x-ray study of phonon broadening and
  charge-density wave formation in ortho-II-ordered
  YBa$_2$Cu$_3$O$_{6.54}$}},}\ }\href@noop {} {\bibfield  {journal} {\bibinfo
  {journal} {Phys. Rev. B}\ }\textbf {\bibinfo {volume} {88}},\ \bibinfo
  {pages} {054506} (\bibinfo {year} {2013})}\BibitemShut {NoStop}%
\bibitem [{\citenamefont {Ishizaka}\ \emph {et~al.}(2004)\citenamefont
  {Ishizaka}, \citenamefont {Arima}, \citenamefont {Murakami}, \citenamefont
  {Kajimoto}, \citenamefont {Yoshizawa}, \citenamefont {Nagaosa},\ and\
  \citenamefont {Tokura}}]{ishi04}%
  \BibitemOpen
  \bibfield  {author} {\bibinfo {author} {\bibfnamefont {K.}~\bibnamefont
  {Ishizaka}}, \bibinfo {author} {\bibfnamefont {T.}~\bibnamefont {Arima}},
  \bibinfo {author} {\bibfnamefont {Y.}~\bibnamefont {Murakami}}, \bibinfo
  {author} {\bibfnamefont {R.}~\bibnamefont {Kajimoto}}, \bibinfo {author}
  {\bibfnamefont {H.}~\bibnamefont {Yoshizawa}}, \bibinfo {author}
  {\bibfnamefont {N.}~\bibnamefont {Nagaosa}}, \ and\ \bibinfo {author}
  {\bibfnamefont {Y.}~\bibnamefont {Tokura}},\ }\bibfield  {title} {\enquote
  {\bibinfo {title} {{Commensurate-Incommensurate Crossover of Charge Stripe in
  La$_{2-x}$Sr$_{x}$NiO$_{4}$ ($x\sim1/3$)}},}\ }\href@noop {} {\bibfield
  {journal} {\bibinfo  {journal} {Phys. Rev. Lett.}\ }\textbf {\bibinfo
  {volume} {92}},\ \bibinfo {pages} {196404} (\bibinfo {year}
  {2004})}\BibitemShut {NoStop}%
\bibitem [{\citenamefont {Daou}\ \emph {et~al.}(2010)\citenamefont {Daou},
  \citenamefont {Chang}, \citenamefont {LeBoeuf}, \citenamefont
  {Cyr-Choiniere}, \citenamefont {Laliberte}, \citenamefont {Doiron-Leyraud},
  \citenamefont {Ramshaw}, \citenamefont {Liang}, \citenamefont {Bonn},
  \citenamefont {Hardy},\ and\ \citenamefont {Taillefer}}]{daou10}%
  \BibitemOpen
  \bibfield  {author} {\bibinfo {author} {\bibfnamefont {R.}~\bibnamefont
  {Daou}}, \bibinfo {author} {\bibfnamefont {J.}~\bibnamefont {Chang}},
  \bibinfo {author} {\bibfnamefont {David}\ \bibnamefont {LeBoeuf}}, \bibinfo
  {author} {\bibfnamefont {Olivier}\ \bibnamefont {Cyr-Choiniere}}, \bibinfo
  {author} {\bibfnamefont {Francis}\ \bibnamefont {Laliberte}}, \bibinfo
  {author} {\bibfnamefont {Nicolas}\ \bibnamefont {Doiron-Leyraud}}, \bibinfo
  {author} {\bibfnamefont {B.~J.}\ \bibnamefont {Ramshaw}}, \bibinfo {author}
  {\bibfnamefont {Ruixing}\ \bibnamefont {Liang}}, \bibinfo {author}
  {\bibfnamefont {D.~A.}\ \bibnamefont {Bonn}}, \bibinfo {author}
  {\bibfnamefont {W.~N.}\ \bibnamefont {Hardy}}, \ and\ \bibinfo {author}
  {\bibfnamefont {Louis}\ \bibnamefont {Taillefer}},\ }\bibfield  {title}
  {\enquote {\bibinfo {title} {{Broken rotational symmetry in the pseudogap
  phase of a high-$T_c$ superconductor}},}\ }\href@noop {} {\bibfield
  {journal} {\bibinfo  {journal} {Nature}\ }\textbf {\bibinfo {volume} {463}},\
  \bibinfo {pages} {519--522} (\bibinfo {year} {2010})}\BibitemShut {NoStop}%
\bibitem [{\citenamefont {Lawler}\ \emph {et~al.}(2010)\citenamefont {Lawler},
  \citenamefont {Fujita}, \citenamefont {Lee}, \citenamefont {Schmidt},
  \citenamefont {Kohsaka}, \citenamefont {Kim}, \citenamefont {Eisaki},
  \citenamefont {Uchida}, \citenamefont {Davis}, \citenamefont {Sethna},\ and\
  \citenamefont {Kim}}]{lawl10}%
  \BibitemOpen
  \bibfield  {author} {\bibinfo {author} {\bibfnamefont {M.~J.}\ \bibnamefont
  {Lawler}}, \bibinfo {author} {\bibfnamefont {K.}~\bibnamefont {Fujita}},
  \bibinfo {author} {\bibfnamefont {Jhinhwan}\ \bibnamefont {Lee}}, \bibinfo
  {author} {\bibfnamefont {A.~R.}\ \bibnamefont {Schmidt}}, \bibinfo {author}
  {\bibfnamefont {Y.}~\bibnamefont {Kohsaka}}, \bibinfo {author} {\bibfnamefont
  {Chung~Koo}\ \bibnamefont {Kim}}, \bibinfo {author} {\bibfnamefont
  {H.}~\bibnamefont {Eisaki}}, \bibinfo {author} {\bibfnamefont
  {S.}~\bibnamefont {Uchida}}, \bibinfo {author} {\bibfnamefont {J.~C.}\
  \bibnamefont {Davis}}, \bibinfo {author} {\bibfnamefont {J.~P.}\ \bibnamefont
  {Sethna}}, \ and\ \bibinfo {author} {\bibfnamefont {Eun-Ah}\ \bibnamefont
  {Kim}},\ }\bibfield  {title} {\enquote {\bibinfo {title} {{Intra-unit-cell
  electronic nematicity of the high-$T_c$ copper-oxide pseudogap states}},}\
  }\href@noop {} {\bibfield  {journal} {\bibinfo  {journal} {Nature}\ }\textbf
  {\bibinfo {volume} {466}},\ \bibinfo {pages} {347--351} (\bibinfo {year}
  {2010})}\BibitemShut {NoStop}%
\bibitem [{\citenamefont {Mesaros}\ \emph {et~al.}(2011)\citenamefont
  {Mesaros}, \citenamefont {Fujita}, \citenamefont {Eisaki}, \citenamefont
  {Uchida}, \citenamefont {Davis}, \citenamefont {Sachdev}, \citenamefont
  {Zaanen}, \citenamefont {Lawler},\ and\ \citenamefont {Kim}}]{mesa11}%
  \BibitemOpen
  \bibfield  {author} {\bibinfo {author} {\bibfnamefont {A.}~\bibnamefont
  {Mesaros}}, \bibinfo {author} {\bibfnamefont {K.}~\bibnamefont {Fujita}},
  \bibinfo {author} {\bibfnamefont {H.}~\bibnamefont {Eisaki}}, \bibinfo
  {author} {\bibfnamefont {S.}~\bibnamefont {Uchida}}, \bibinfo {author}
  {\bibfnamefont {J.~C.}\ \bibnamefont {Davis}}, \bibinfo {author}
  {\bibfnamefont {S.}~\bibnamefont {Sachdev}}, \bibinfo {author} {\bibfnamefont
  {J.}~\bibnamefont {Zaanen}}, \bibinfo {author} {\bibfnamefont {M.~J.}\
  \bibnamefont {Lawler}}, \ and\ \bibinfo {author} {\bibfnamefont {Eun-Ah}\
  \bibnamefont {Kim}},\ }\bibfield  {title} {\enquote {\bibinfo {title}
  {{Topological Defects Coupling Smectic Modulations to Intra--Unit-Cell
  Nematicity in Cuprates}},}\ }\href@noop {} {\bibfield  {journal} {\bibinfo
  {journal} {Science}\ }\textbf {\bibinfo {volume} {333}},\ \bibinfo {pages}
  {426--430} (\bibinfo {year} {2011})}\BibitemShut {NoStop}%
\end{thebibliography}%
\newpage
\setcounter{figure}{0}
\def\thefigure{S\arabic{figure}}

\noindent{\bf Supplemental Material}
\bigskip

\begin{figure*}[b]
 \centering
    \includegraphics[width=1.6\columnwidth]{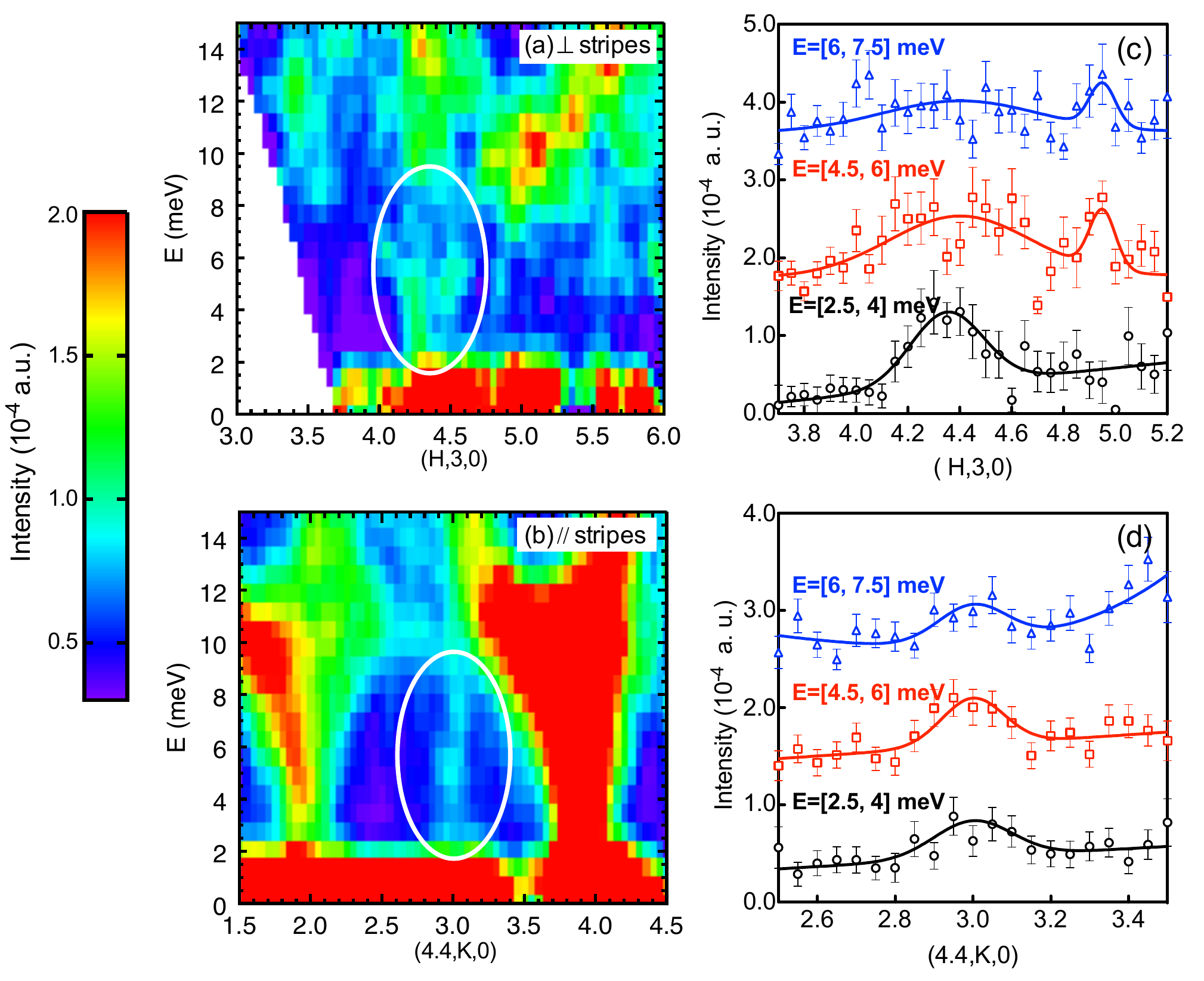}
    \caption{\label{fig:FigS1} (color online)  Low-energy excitations of charge stripes in \LSNOn\ at 220~K. (a) Scattered intensity as a function of $E$ vs \textbf{Q} through the charge-order position ${\bf Q}^*_{\rm co}=(4.44, 3, 0)$ along the [100] direction, with integration over $2.9\leq K\leq3.1$; (b) similar to (a) but along [010] and integrated over $H\pm0.3$. (c) Cuts through (a), integrated over 1.5-meV in energy, as noted in the labels, and shifted vertically for clarity. (d) Cuts through (b).  Lines in (c) and (d) are least-squares fits to Gaussian peaks plus a background.  The extra peak in (c) near (5,3,0) is a spurious feature, as discussed in the text.  In all cases, the data have been integrated over 
$-0.2\leq L\leq0.2$, and a {\bf Q}-independent incoherent elastic scattering contribution, broadened by instrumental energy resolution, has been subtracted.}
\end{figure*}

Here we present further results that may be of interest to experts.  Regarding the experimental measurements at HYSPEC, the incident energy was fixed at 50~meV, with a Fermi chopper frequency of 360~Hz. Data were collected while rotating the sample about the vertical axis in steps of 0.5$^{\circ}$ to map a volume in reciprocal space.  The detector, which covers $60^\circ$ of scattering angle, was positioned with its center at 35$^{\circ}$ or 90$^{\circ}$ to cover either small or large wave vector (\textbf{Q}) ranges.

Measurements of the charge stripe fluctuations obtained at $T=220$~K are shown in Fig.~\ref{fig:FigS1}.  These results are similar to those obtained at 160 K (see Fig.~4 of the main paper).  The charge stripe fluctuations observed at 220~K are slightly more intense, and somewhat broader in $H$ and $K$, than those at 160~K, but the anisotropy of the dispersion is still clear.

\begin{figure}[t]
 \centering
    \includegraphics[width=0.9\columnwidth]{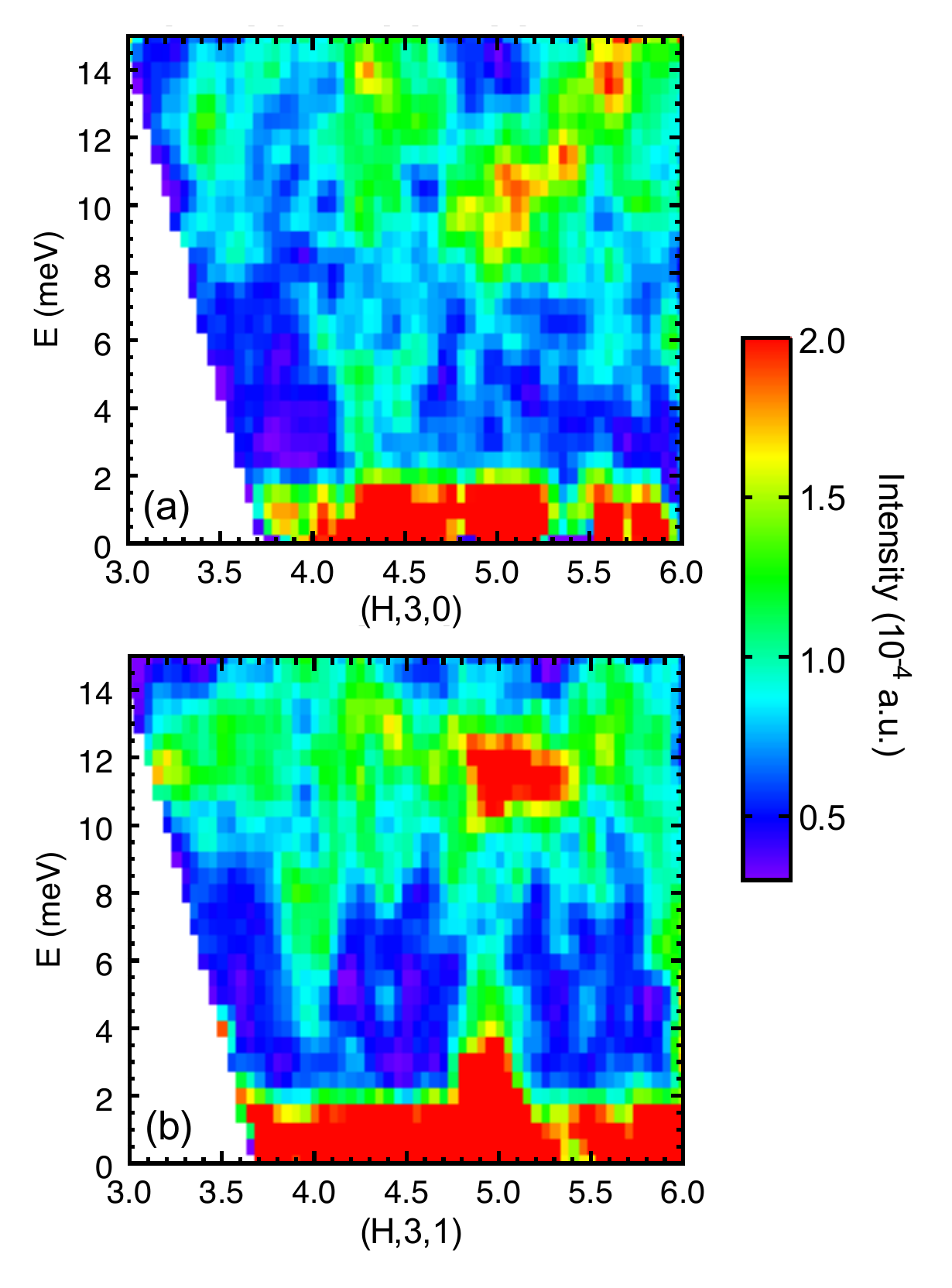}
    \caption{\label{fig:FigS2} (color online) Data slices at 220~K along ${\bf Q}= (H, 3, 0)$ (a) and $(H, 3, 1)$ (b), with integration over $2.9\leq K\leq3.1$, $L_0\pm 0.2$.}
\end{figure}

In Fig.~\ref{fig:FigS1}(a) and (c), there is a spurious feature near $(5,3,0)$ for energies between 4 and 7 meV.  This feature might be due to noise, but it could also be associated with the tail of phonons dispersing about the allowed Bragg peak at $(5,3,1)$.  To illustrate this issue, we compare in Fig.~\ref{fig:FigS2} data slices along ${\bf Q}=(H,3,0)$ [the same as Fig.~\ref{fig:FigS1}(a)] and along $(H,3,1)$.  Besides the acoustic phonons at (5,3,1), one can also see soft phonons, associated with tilting modes of the NiO$_6$ octahedra, at (4,3,1) and (6,3,1). Keep in mind that we have coarse resolution (due to vertical focusing) in the [001] direction and that the phonon dispersion is generally weaker in that direction.  The charge fluctuation scattering is definitely broad in $L$; however, we have to integrate over a restricted $L$ range in order to minimize the contamination from various lattice excitations with minima at $L=\pm1$.

\end{document}